\documentclass[12pt,preprint]{aastex}
\slugcomment{Accepted and scheduled for publication  
in {\it The Astrophysical Journal},  for  the August 10, 2011, v 737, 1st issue}  
\def\lax {\ifmmode{_<\atop^{\sim}}\else{${_<\atop^{\sim}}$}\fi}  
\def\gax {\ifmmode{_>\atop^{\sim}}\else{${_>\atop^{\sim}}$}\fi}  
\def\gtorder{\mathrel{\raise.3ex\hbox{$>$}\mkern-14mu
             \lower0.6ex\hbox{$\sim$}}}

\def\qcor{Q_{\rm cor}}
\def\qtot{Q_{\rm tot}}

\def\qd{Q_{\rm disk}}

\def\cm2{cm$^{-2}$}
\def\s1{s$^{-1}$}

\def\kte{kT_{\rm e}}

\begin{document}

\title{On the Constancy of the Photon Index of  X-ray spectra of 4U~1728-34 through all spectral states} 
%during outburst transitions}

\author{ Elena Seifina\altaffilmark{1} and Lev Titarchuk\altaffilmark{2}}
\altaffiltext{1}{Moscow State University/Sternberg Astronomical Institute, Universitetsky 
Prospect 13, Moscow, 119992, Russia; seif@sai.msu.ru}
\altaffiltext{2}{Dipartimento di Fisica, Universit\`a di Ferrara, Via Saragat 1, I-44100 Ferrara, Italy, email:titarchuk@fe.infn.it; ICRANET, Piazza della Repubblica 10-12
65122 Pescara,  Italy; George Mason University Fairfax, VA 22030;   
Goddard Space Flight Center, NASA,  code 663, Greenbelt  
MD 20770, USA; email:lev@milkyway.gsfc.nasa.gov, USA}

\begin{abstract}
%We investigate the evolution of spectral and timing properties in a number of 
%Galactic black hole (BH) sources during spectral transitions.
% in order
%to study the implications of scaling of a correlation between  photon index of Comptonized component 
%and the  centroid  of  low frequency quasi-periodic  oscillations (QPO) in the X-ray energy spectrum. 
%We also  analyze  the evolution  of Fourier Power Density Spectra along with the energy spectral evolution. 
%Particularly we are  interested in the behavior  
%of low-frequency quasi-periodic  oscillations (QPOs) vs photon index (spectral state).
%We present the study of correlations between spectral  of timing 
We present an analysis of the spectral properties observed in X-rays from 
Neutron Star X-ray binary 4U~1728-34 during 
%{\it  burst}
 transitions between the low and the high luminosity states when  electron temperature $kT_e$ of the  Compton cloud monotonically decreases  from 15 to 2.5  keV.  We analyze 
 %quite a few   
 the  transition episodes 
from this source  observed with {\it Beppo}SAX  and  {\it Rossi} X-ray Timing Explorer
({\it RXTE})  satellites. 
  We find that the X-ray broad-band energy spectra of 4U~1728-34 during all spectral states can be  modeled by  a combination of a thermal (black body-like) component, a Comptonized 
component (which we herein denote {\it COMPTB}) and 
a {\it Gaussian} component. 
Spectral analysis using  this model evidences that  the photon power-law index $\Gamma$ is almost constant ($\Gamma=1.99\pm 0.02$) when   $kT_e$ changes  from 15 to 2.5 keV during these spectral transitions.
% between the low and high luminosity states respectively. 
%As a consequence 
%of quasi-constancy of $\Gamma$ around mean value of 1.97$\pm$0.01, we described (modelled) burst 
%evolution of 4U~1728-34 in terms of the changes (variations) of the electron temperature $kT_e$ of 
%Comptonized region. In this way, transition of 4U~1728-34 from ``low luminosity state'' to 
%``high luminosity state'' corresponds tochanges (decreasing) of the electron temperature from 15 to 2.5 keV.
%}
% We also present an observable correlation between photon index and the {\it COMPTB} normalization. 
%In frame of this model 
%Using  the disk ``seed'' ({\it COMPTB}) normalization, which is 
%proportional to the total (sub-Keplerian plus  disk) mass accretion rate, we found the constancy of  photon index $\Gamma$ at a value   of 2
%is fundamental to establish the index stability effect 
%during the transition from low to high luminosity state. %, although during these 
%observations the normalization electron temperature of the Compton cloud changes by factor 3.
% We do find the stability  
%of the photon index of the Comptonized component versus the {\it COMPTB} normalization at a value of 2 
%during all the spectral  states. 
We 
%discussed a physical model (Farinelli \& Titarchuk, 2011)  
explain  this quasi-stability of the index $\Gamma$ 
%using the Farinelli \& Titarchuk spectral model.   In the framework of
by  the  model  in which the spectrum  is  dominated 
by the strong thermal Comptonized component formed  in the transition layer (TL) between the  accretion disk and neutron star surface. 
%In the framework of this model we  find  that 
The index quasi-stability 
%(small variability)  
takes place when the  energy release in the TL is much higher than the flux  coming to the TL from the accretion disk. Moreover, this index stability effect now established for  4U~1728-34 
during spectral evolution of the source 
%from the low to high luminosity states 
was previously  suggested for a number  of other neutron binaries (see Farinelli \& Titarchuk, 2011).
 This   intrinsic property of  neutron star   is fundamentally  different   from that in black hole binary sources for which the index monotonically increases during spectral transition from the low state  to high state and saturates at high values of mass accretion rate.
%We argue that the index stability effect is due 
%to dominatinf of Comptonization in the Transition Layers close to NS and
%therefore propose it as a signature of NS.

\end{abstract}


\section{Introduction}

The evolution of spectral parameters of compact objects in X-ray binaries is 
of great interest for understanding the nature of compact objects. It is well 
known that  a  number of black hole (BH) candidate sources demonstrate correlations 
between their 1-10 Hz quasi-periodic oscillation frequencies  (QPOs)  $\nu_L$ and photon
power-law index $\Gamma$  during  spectral transition when sources evolve from the low  to high
 states [see  \cite{st09}, hereafter ST09].  Then  the definition of the spectral state is related to the level of  soft blackbody emission presumably related to mass accretion rate.  In the high states of BHs these index-QPO frequency 
correlations {\it sometimes}  show a saturation  of $\Gamma$  at high values of  $\nu_L$. 
On the other hand  ST09 [see also \cite{tsei09}, hereafter TS09]  found that  $\Gamma$  saturates 
with mass accretion rate in almost  any case of a BH binary. This saturation effect can be  considered   as a black hole signature or signature of the converging flow into BH (ST09 and TS09).  The question naturally arises how the spectral index behaves as  a function of mass accretion rate or as a function of cutoff energy of the spectrum in neutron star (NS) sources.  
%{\it 
%However, is it posiible to indicate the like effect, which are typical of neutron star?
%}
  
Recently  \cite{ft10}, hereafter FT11, collected  X-ray spectra obtained by {\it Beppo}SAX   
 from quite a few NS sources Sco~X-1, GX~17+2, Cyg~X-2, GX~340+0, GX~3+1 and GS~1826-238.   Their results probably  indicate   that  the value of  the photon index slightly varies around 2 independently of the spectral state 
 (or electron temperature of Compton cloud) at least for this particular sample of  NS  spectra   
 [see \cite{disalvo2000a}; \cite{F08}]. 
 However, the available data for those sources were taken when these sources  were  in the high state or in the low state  but nobody has  analyzed  up to now the spectral evolution from the low state to the high state for  some particular  NS source.  

A  suitable candidate for the study of the spectral evolution in NS is 
the so called atoll 4U~1728-34 which  
exhibits  a remarkable  spectral transition from the  low state to the high state and vice versa.
%frequent bursts characteristic. 
4U~1728-34 (GX~354$-$0) was  first resolved by {\it UHURU} scans of the Galactic center region in 
1976 [see \cite{form76} and \cite{bradt93}]. Then type I X-ray bursts from 4U~1728-34 were discovered 
during SAS-3 observations by {\cite{Lew76} and  \cite{hof76}. Further the
bursting behavior was subsequently studied in detail using
extensive observations by SAS-3, which accumulated
96 bursts in total. Using  these data \cite{bas84} 
presented evidence for a narrow distribution of peak burst
fluxes, as well as a correlation between the peak flux and
the burst phase. The distance to the source in the range of 4.2 -- 6.4 kpc has been estimated 
by \cite{vanp78} and confirmed by \cite{bas84} and \cite{kam89} using  measurements of the 
peak burst fluxes.   
%(van Paradijs 1978;
%Basinska et al. 1984; Kaminker et al. 1989). 

A radio counterpart of 4U~1728-34 was detected during VLA (at 4.86 GHz) observations
with a variable flux density in the range of $\sim$0.3 -- 0.6 mJy~\citep{marti98}.
% presented observations of  4U~1728 in the radio band (at 4.86 GHz) 
%with VLA when they were able  to successfully detect its radio counterpart with a variable flux 
%density in the range $\sim$0.3 -- 0.6 mJy.
 The estimated
extinction of the source is $A_V\simeq$14 and a precise
position following from the detection of the radio counterpart
allowed to  identify  this source as a K = 15 infrared source~\citep{marti98}.
%(Marti et al. 1998). 
%No independent distance measurement is available. 
Long-term {\it Ariel-5} measurements, as well as
the extensive monitoring by the All-Sky Monitor
(ASM) on the board of {\it RXTE}, suggest the presence of a long-term
quasi-periodicity  about   63-72 d  \citep{kong98}.
%(Kong et al. 1998).
{\it RXTE}/PCA observations of the source in 1996 led to
the discovery of nearly coherent millisecond oscillations
during the X-ray bursts \citep{str96}.  
%(Strohmayer et al. 1996). 
%%%%%BEGIN

\cite{to99}
%Titarchuk, \& Osherovich (1999)
presented a model for the radial oscillations and diffusion of the perturbation 
in the transition layer  (TL) surrounding the neutron
star.  Using dimensional analysis, they  have identified the corresponding
radial oscillation and diffusion frequencies in the TL with the low-Lorentzian $\nu_L$ and break
frequencies $\nu_b$ for 4U~1728-34. They predicted values for $\nu_b$ related to the diffusion in the boundary layer, that are consistent with the observed $\nu_b$.
%%%%END

Subsets of the bursts observed during the PCA observations
have been also studied   by \cite{vanst01}
%traaten et al. (2001) 
and \cite{franco01}
with particular attention to the relationship between
the appearance of burst oscillations and a value of mass
accretion rate. 

\cite{ts05}, hereafter TS05, analyzed {\it RXTE}/PCA observations of  4U~1728-34 
in energy range from 3 keV to 40 keV and they found that using the model comprising 
 two Comptonization components (BMC
\footnote{BMC is so called ``Bulk Motion Comptonization'' {\tt XSPEC} Model 
(see detailes in section 6.2.10 of ``User's Guide of an X-Ray Spectral Fitting Package XSPEC v.12.6.0'' 
and http://heasarc.gsfc.nasa.gov/xanadu/xspec/manual/Additive.html)}
)   the photon index $\Gamma$ is consistent 
with being  quasi-constant  (around 2.2), while  the low quasi-periodic oscillation (QPO)  
frequency does not exceed  10 Hz,  then $\Gamma$    monotonically increases   
to  values of 6.  Moreover, using 
broad-band observations of 4U~1728-34 by $Beppo$SAX \cite{disalvo2000a} and \cite{piraino00} 
fitted  the X-ray spectra of 4U~1728-34 in a wide energy range from 0.1 keV to 200 keV 
by a sum of $Blackbody$ component plus a thermal Comptonization spectrum, usually described 
by the XSPEC COMPTT model~[\cite{tl94}, \cite{HT95}]. 
%Furthermore,                                                              
%{\it Beppo}SAX observations in a wide energy 
%range from 0.1 keV to 200 keV,  
%Also FT11  found that X-ray spectra can 
%also 
%}
% be 
%[[only]] fitted by  
%a model  consisting  one Comptonized component ({\it COMPTB}) plus low-energy 
%blackbody-like  component ({\it bbody}). 
The TS05 model cannot fit 
the {Beppo}SAX data and moreover this  model has more parameters than that of 
 model of \cite{disalvo2000a}, \cite{piraino00} and 
FT11. Naturally  we pose the  question if  the  FT11  model can fit the 
{\it Beppo}SAX  along with {\it RXTE} data and what kind of the dependence 
of index  vs spectral state can be found.      
 
% and higher on the base of  RXTE burst observations (1996 -- 2001). 
%They revealed the spectral evolution of the
%Comptonized blackbody spectra when the source transits from the hard to soft
%states. It was provided that the hard state spectrum has a typical thermal Comptonization spectrum
%of the soft photons which originated in the disk and the NS outer photospheric
%layers. They also stated that the hard state photon index is $\Gamma\sim$ 2 
%while the soft state spectrum consists of two blackbody components which are 
%only slightly Comptonized.
%Recently  Farinelli and Titarchuk (2010) showed that spectra of 4U~1728-34 as well as the numbers of NS sources 
%can be fitted by $CompTB$ model (Farinelly et al., 2008), wherein spectral index $\alpha$ is around 1.

%As mentioned above,
%the observed spectral index $\alpha$ for the 
%numbers of NS sources, in particular for 4U~1728-34 based on $BeppoSAX$ observations, lies in a belt around 1$\pm$0.2.

 In this Paper we do  try to answer a fundamental question 
 %of high energy astrophysics 
 on the possibility  to distinguish BH from NS systems, which extensively attempted to be solved 
without considering the mass of the compact object as a main argument, in particular without using optical 
counterpart data to measure mass function. In this way some methods have been 
proposed to identify systems contains BHs using X-ray observational properties only. 
The strong rapid variability was firstly considered as a signature of the presence of BH (Oda et al., 1971),  until the same rapid variability was detected  in accreting NSs \citep{tennant86}. 
%(Tennant et al., 1986) 
Now it is well established  
%it was detected 
that Galactic BH candidates (BHCs)  demonstrate two spectral states, 
the ``high state'' (HS) and ``low state'' (LS)  and transition between them 
%during burst transitions~
\citep{RM06}. However, sometimes so 
called $atoll$-NS sources \footnote{Here, we use a term  of the  $atoll$-NS sources   to specify  NS X-ray binaries characterized by   a specific ``$\epsilon$''-shaped track in color-color diagram.
%, based on its X-ray spectral properties. 
}  also show  the ``high'' and ``low'' spectral states 
[\cite{dai2006}, TS05]. Therefore this property requires more detailed investigations  
BHs vs NSs. Specifically, the HS spectra  of BHCs are characterized  by thermal emission at $\sim$1 keV 
presumably originated in  the  accretion  disk, along with a steep power law tail whose  photon index 
($\Gamma$=2$-$3) monotonically increases with mass accretion rate (see ST09).
While the LS spectra  show  much weaker disk emission than that in the  HS spectra and  harder power-law 
tail (which  photon index is around $\Gamma\sim$~1.7). 
This hard component generally believed is   a result of  thermal Comptonization of soft (disk) photons in 
a hot gas (Compton cloud) in vicinity of the compact object.
%%%%BEGIN

It is worth to point out  that BHs, in  contrast with NSs, sometimes demonstrate more complicate X-ray spectrum. For example, the  {\it RXTE} %X-ray
 spectra of BH GRS~1915+105   require two  Comptonization components, soft and hard ones [see \cite{tsei09}].   
%This case arises in  BH GRS~1915+105 where 
In this case one can clear see the evolution of two   photon indices $\Gamma_1=1.7 - 3.0$  and 
$\Gamma_2=2.7 - 4.2$
for  the hard and soft components respectively. 
% , for the hard  and soft $\Gamma_2=2.7 -- 4.2$. 
%components during spectral state transition~(TS09). 
\cite{tsei09} argued that the index saturation effect  of the hard component is due to Comptonization of   the soft (disk)  photons in the converging flow into BH 
and  that of the soft component is due to the thermal Comptonization   in the transition layer when mass accretion rate  increases. These conclusions were later supported by  Monte-Carlo simulations by \cite{lt10}.

%%%END
 Moreover,  in the description of  BH and NS X-ray low state spectra with the thermal 
Comptonization model, there is essential difference between these types of the compact sources.   The electron temperature of the Compton (scattering)  cloud $kT_e$ is usually  lower for  NSs, $kT_e<25$ keV  than that for BHs, $kT_e>50$ keV 
[see 
%Churazov et al., 1997
\cite{churaz97}]. The lower electron temperature in NSs than that in BHs is   a consequence of the 
additional cooling provided by the NS surface which  reflects  X-ray photons and ultimately determines  a value of the CC electron temperature  [\cite{tlm98} and see also \cite{ST89}, \cite{kluzh93}]. 
%%%%BEGIN

This  fact of  the observed difference between the black hole and neutron stars 
%in quiescent  
 was  recently also discussed by \cite{RM11}. 
 %Reynolds \& Miller (2011) 
% and based on revealed fact that the black hole systems  usually  be fainter (Garcia et al. 2001) in comparison of the quiescent luminosities of the neutron star binaries.  
They conclude  the observable difference of $kT_e$ in these types of the sources  is 
 as an evidence  of the absence and presence  of a solid surface in BHs  and NSs respectively   and this fact can be considered as an indirect evidence for the  existence of the event horizon in BHs.   It is worth to point out  that \cite{tlm98} and \cite{tf04} previously came to the similar conclusions based on the analysis of the Compton cooling of X-ray emission region (transition layer) in the presence (NS) and absence (BH) of the reflection  surface. 

Thus,  the basic property, which  differs BHs  from NSs, is the presence of event 
horizon as well as a converging flow in vicinity of BH \citep{ebisawa96}.
%(e.g., Ebisawa, Titarchuk \& Chakrabarti 1996).
%Moreover, to describe spectra during hard  states of BH binaries (LHS and IS) a significant emission 
%in the hard X-rays/soft $\gamma$-ray range should be accounted. For this reason the steep power-law 
%was interpreted in terms of Comptonization in a converging bulk flow in the vicinity of the BH 
%(e.g., Ebisawa, Titarchuk \& Chakrabarti 1996).
 In fact, close to the event horizon, 
the strong gravitational force dominates the pressure forces which leads to an almost  free fall 
converging flow of  accreting material into a  BH. The  dynamical Comptonization of low energy photons  
off  fast-moving electrons dominates the thermal Comptonization at high mass accretion rate, then the 
plasma temperature of converging flow  is less that 10$-$15 keV,  and as a result  an extended steep 
power law is formed  [see \cite{tmk97}, \cite{tz98} and \cite{lt99}].
%Titarchuk et al., 1997; Titarchuk \& Zannias 1998 \cite{lt99}].
%with photon index $\sim$2.5 mostly 
This kind of the spectra are observed in the soft state of BH binaries (see e.g. ST09 and  TS09).  

On the other hand in NS sources  the radiation pressure forces become 
dominant close to their surface  and thus a free fall  should be suppressed at high mass accretion rates.  
Does the presence of the firm surface in NS makes any difference as for the dependence of the photon index 
$\Gamma$ vs mass accretion rate with respect to that established in BHs (see ST09, TS09 and ST10)? Furthermore 
can  the index saturation detected in many BHs with mass accretion rate  exist only in BH sources and cannot  be  observed so far   in NS sources? 
For example \cite{disalvo2006},  studying of low mass X-ray binaries hosting neutron stars, concluded that it is unlikely to distinguish BHs from  NSs based on their X-ray spectra.
%except for the fact that black holes system show sometimes a  more extreme behavior with respect to neutron star systems. 
 However FT11 argue that in NS sources the index $\Gamma$ weakly  varies until the soft photon illumination of the transition layer  (TL)  $Q_d$ is much smaller than the energy release  
in the TL, $Q_{cor}$.    
We  try further to test this kind of the index behavior in the  NS source  using X-ray observations  of  the  $atoll$  source 4U~1728-34  and  compare it,  
if it's possible,  with the index dependence on mass accretion rate established  in BHs.
In this Paper we present the analysis of the {\it Beppo}SAX  available observations 
during  1998 -- 1999 years and  {\it RXTE}/PCA/HEXTE observations during  1996 - 2000 years for 
4U~1728-34.  In \S 2 we present the list of observations used in our data analysis while 
in \S 3 we provide the details of X-ray spectral analysis.  We analyze an evolution of 
X-ray spectral and timing  properties during the state transition in \S 4 and \S 5.  
We discuss our results and make our  conclusions in \S 6 and \S 7. 
 
\section{Data Selection \label{data}}

Broad band energy spectra of the source were obtained
combining data from  three {\it BeppoSAX} Narrow
Field Instruments (NFIs): the Low Energy Concentrator
Spectrometer [LECS; \cite{parmar97}] for 0.3 -- 4 keV, the Medium Energy Concentrator Spectrometer
[MECS; \cite{boel97}] for 1.8$-$10 keV and the Phoswich Detection
System [PDS; \cite{fron97}] for 15$-$ 60 keV. 
 The SAXDAS data analysis package is used for processing data. 
For each of the instruments we perform the spectral analysis in the energy range 
for which response matrix is well determined. The LECS data have been 
re-normalized based on MECS. Relative normalization of the NFIs were treated 
as free parameters in
 model fitting, except for the MECS normalization that was fixed at a value
 of 1. We checked after that the fitting procedure if these normalizations
 were in a standard range for each
 instruments% (section 4.2 of Cookbook for the BeppoSAX NFI spectral analysis
\footnote{http://heasarc.nasa.gov/docs/sax/abc/saxabc/saxabc.html}.

 Specifically, LECS/MECS re-normalization ratio is 0.92 and PDS/MECS
 re-normalization ratio is 0.97. 
In addition,  spectra are rebinned  accordingly to  
 energy resolution of the instruments in order to obtain
 % independent 
significant data points.  We rebinned  the LECS spectra applying a rebinning template for grouping
 (lecs\_2.grouping)   with  an energy depending  binning factor  used  
%(Sect.3.1.6 of Cookbook for the BeppoSAX NFI spectral analysis) using
% rebinnig template files
 in GRPPHA of
 XSPEC \footnote{http://heasarc.gsfc.nasa.gov/FTP/sax/cal/responses/grouping}. Also we rebinned  the PDS spectra with linear binning
 factor 2, grouping two bins together (resulting bin width is 1 keV).
In Table 1 we list  the {\it Beppo}SAX observations used in our analysis. 

We also  analyzed the  available data obtained with {\it RXTE} ~\citep{bradt93} which have been 
%(Bradt et. al, 1993) 
found  in the  time period from February 1996 to July 2000 [see also  a review by  \cite{gall08}]. 
%In general, there are 106 bursts from
%this source in the {\it RXTE} burst catalogue \citep{gall08}. 
In our investigation we selected 
127 observations made at different count rates (luminosity states)
%}
% states 
%{\it (luminosity) }states of 4U~1728-34 
with a good coverage of rise-decay flare track. 
%%In general we investigate an evolution of X-ray emission from 4U 1728-34 during 50128-51733 MJD interval 
% {piraino00} 
%%within 
%one continuos  cycle of mean ASM flux variability (see Fig. \ref{outburst_05_rise}).
%{\it
We have  made an analysis of {\it RXTE} observations  of 4U~1728-34  during four years 
for 8 intervals indicated by  blue rectangles in Figure~\ref{outburst_05_rise} ($top$). 
We have also  analyzed two {\it Beppo}SAX observations which dates  marked by green triangles there.
%Figure~\ref{outburst_05_rise}.  
%and   \ref{evolution_lc_3}).
%}
 
Standard tasks of the HEASOFT/FTOOLS
5.3 software package were utilized for data processing.
For spectral analysis we used PCA {\it Standard 2} mode data, collected 
in the 3 -- 20~keV energy range. The standard dead time correction procedure 
has been applied to the data. 
% BEGIN
%{\it 
The average dead time  correction is in the range  3 -- 10 \% depending on the count rate value.
%}
%END

HEXTE data have been also  used in order to construct broad-band spectra.
We  subtracted the background corrected  in  off-source observations. 
%In order to account for the uncertainties in the HEXTE response and background determination
To exclude the channels with largest uncertainties
we used  only data  in  20 -- 60~keV energy range 
%were used 
for the spectral analysis. 
The HEXTE data have been re-normalized based on the PCA. 
% BEGIN
%{\it 
Typical PCA/HEXTE renormalization factor is 0.98.
%}
%END
We used the data which  are available through the GSFC public archive 
(http://heasarc.gsfc.nasa.gov). 
In Table 2 we list    the groups
 of {\it RXTE} observations which cover  
%complete dynamical range {\it LHS-(IS)-LHS }
 the source evolution  from  quiescent   to flare events. 
%We present here period ranges MJD=50462 -- 51081 and MJD=50868 -- 54096, as set 
%of intervals of 4U~1728-34 quite and outburst transitions between {\it hard} and {\it intermediate} states. 
% BEGIN
%{\it 

Note that we did not use any normalization factor to normalize between BeppoSAX and {\it  RXTE} data.
%}
%END
We also used public 4U~1728-34 data from the  All-Sky Monitor (ASM) 
on-board \textit{RXTE}. 
%\citep{sw99}. 
We retrieved  the ASM light curves (in 2$-$12 keV energy range) 
from the public \textit{RXTE}/ASM archive in MIT
\footnote{http://xte.mit.edu/ASM\_lc.html}. 
%% BEGIN
%{\it
In the $bottom$ panel of  Figure~\ref{outburst_05_rise} we  show a mean count rate (blue dashed line)  during 1996$-$2010 interval of ASM/RXTE monitoring observations of 4U~1728-34. 
In this panel one can also  see a  long-term quasi-periodic variability of mean soft flux 
during  $\sim$ six years cycle. We investigate 
%one (indicated by green 
%double arrow) of 
available periods of slow variability  (indicated by green 
%double arrow)  
during which we also  have 
%chosen as that interval, 
%during which we also have 
 the  $Beppo$SAX observations of 4U~1728-34 (see also  the upper panel of 
 Figure \ref{outburst_05_rise}). 
%}
%% END

%the High Pressure
%Gas Scintillation Proportional Counter (HPGSPC;Manzo et al. 1997) for 8 -- 40 keV, 
%Hereinafter we use terms  ``low'' and ``high'' states to describe evolution of X-ray  spectra during 
%spectral state transitions of NS binary and relate them with the NS luminosity. 
We use definitions of the  low and high luminosity states to relate these states to the source luminosity and we demonstrate that during the high-low state transition the electron temperature of Compton cloud changes from 2.5 keV to 15 keV respectively and vice versa.  
Thus the ``high spectral state'' corresponds to ``low electron temperature state'' and vice versa
the ``low spectral state'' corresponds to ``high electron temperature state''.  
During the  flare, seen  in the ASM light curve, the electron temperature $kT_e$ usually decreases from 15 keV   to 2.5 keV.
%Usually,  ``hard'' and ``soft'' states  (of BHs and sometimes for NSs) are related to the index of the power law part of the BH spectra. But for NSs the index is mainly low variable (except for $Z$ and $atoll$ sources), moreover as it will be shown (demonstrated) below, is 
%independent of the luminosity and stays almost the same when electron temperature of Comptoncloud changes from 2.5 keV to 20 keV. 
%Whereas gradual transition from soft to hard state has been observed rather in response to a decrease of the source 
%X-ray luminosity, this transition is modeled in terms of a gradual decrease of electron temperature $kT_e$ of the  Comptonisation region. Therefore hereinafter we'll refer (use) to ``high luminosity state'' and 
%``low luminosity state'' or simply ``high state'' and ``low state''.
%Therefore, here we refer (denote, define) to 
%``high'' and ``low'' states of NS binary based on higher and lower luminosity during so called 
%burst and quiet states. 
We introduce  a definition of  a ``burst'' to point out   a significant increase of X-ray flux  (about 
factor of  5)  with  respect  to the persistent emission level. 
Specifically,  we call the ``burst'' 
% for 4U~1728-34 is presented (we suppose) 
%considering the start/end of each outburst, 
when ASM count rate is greater than 10 c/s. 
%BEGIN
%{\it 
%during some days 
We associate  the count rate increase  with the increase of mass accretion rate. 
%}
%END
%In the following it will be also shown that
%Get ahead of detailed data analysis we point out that
% ``high spectral state'' corresponds 
%to ``low electron temperature state'' and
%``low spectral state'' corresponds to ``high electron temperature state'' in adopted terms.

\section{Spectral Analysis \label{spectral analysis}}

%It is well known that analysis of RXTE data is characterized by dualizm of solution when 
%compact object transits to soft state. In this respect BeppoSAX observation allow  to choose 
%real solution. 
%Thus two 
%%BeppoSAX observations, which obtained at different active states (low-hard (20889003) and 
%%high-soft (20674001)), provide possibility to understand the tendency of prameter evolution 
%%during outburst based on wide energy range of available BeppoSAX spectra (0.3-100 keV). 
%%In turn extensive RXTE observations 
In our spectral  data analysis we use a model which consists a  sum of Comptonization
 ($COMPTB$) component, 
 % which is characterized by
%temperature of the seed photons $T_s$,
%energy index of the Comptonization spectrum $\alpha$ ($\alpha=\Gamma-1$) and 
%electron temperature $T_e$
 [{\it COMPTB} is the XSPEC Contributed model\footnote{http://heasarc.gsfc.nasa.gov/docs/software/lheasoft/xanadu/xspec/models/comptb.html},
see \cite{F08}, hereafter F08], a soft blackbody component of temperature 
 $T_{BB}$ and Gaussian line component.
% More specifically, %Thus (in this way),
%In particular
  The $COMPTB$ spectral component has the following parameters:
temperature of the seed photons $T_s$,
energy index of the Comptonization spectrum $\alpha$ ($=\Gamma-1$), 
electron temperature $T_e$,   Comptonization  fraction $f$ [$f=A/(1+A)$]
and the normalization of the seed photon spectrum $N_{COMPTB}$ (see Appendix A for the definition of
$N_{COMPTB}$) . 
%We  find that  an addition of  the soft blackbody-like 
%component  of  temperature $T_{BB}=$0.5$-$0.7 keV to the model  significantly improves  the fit quality. 
 
 % (see Table 3 \& 4 for the values of the best-fit parameters).  

In Figure~\ref{geometry}  we illustrate  our spectral model 
%by schematic  view 
as a basic model for  fitting  the {\it Beppo}SAX and {\it RXTE}  spectral data 
for 4U~1728-34.
% we present a considered accretion scenario, which we suggest taking place 
%in the innermost part of a neutron star source. 
We assume that accretion onto a neutron star takes place when the material passing through 
the  main two regions,  a geometrically thin accretion disk [standard Shakura-Sunyaev 
disk, see \cite{ss73}]
%Shakura \& Sunyaev (1973)] 
and the  transition layer (TL), where NS and  disk soft photons   are  upscattered off hot electrons. 
%region. 
In other words, in our picture, the emergent thermal Comptonization spectrum is  formed in the  TL region, where disk  BB-like seed photons and neutron star soft photons are upscattered in the relatively hot plasma. Some fraction of these seed soft photons can be also seen directly by the Earth observer. Red and blue photon trajectories shown in Fig. \ref{geometry} correspond to soft 
(seed) and hard (up-scattered) photons respectively. 

%also interact with the surface of neutron star 
%and/or seen  

%The spectral samples are presented 
%and   
We show examples of X-ray spectra in  Fig.~\ref{BeppoSAX_spectra} (for $Beppo$SAX data)
and in Figs. \ref{rxte_hard_state_spectrum}$-$\ref{rxte_soft_state_spectrum} (for   {\it RXTE}  data). 
Spectral analysis of {\it Beppo}SAX and  {\it RXTE}  observations indicates  that X-ray  spectra of 4U~1728-34 can be   
described by the model where its Comptonization component can be presented by  $COMPTB$ model. %(Farinelli et al, 2008)
% at high energies. 
Moreover,  for  broad-band {\it Beppo}SAX  observations this spectral model  component is  modified  by  
photoelectric absorption at low energies. 
% was  evidently rejected, $\chi^2_\nu=$2.15 for 448 d.o.f.  It is characterized by strong residuals at photon enrgies between 1 and 4 keV. 
Also  following  to \cite{disalvo2000a} and \cite{piraino00} suggestions, 
we add  $Gaussian$ line at $\sim$6.7 keV and thermal blackbody  component at low energies (1 -- 4 keV)  to improve 
the  fit statistics. Note along with these components    \cite{disalvo2000a}  included a narrow $Gaussian$ line  
to fit an excess in the residuals around 1.7 keV. We also test  the presence of this line feature, but  the addition 
of this component to the model does not improve a quality of the model fit.  
It is worth noting  that \cite{dai2006} analyzed   a simultaneous $Chandra$ and {\it RXTE} observations of the 4U~1728-34  (2002 March, 2 -- 5).  
They  fitted the 1.2 -- 35 keV continuum spectrum with a blackbody plus a Comptonized component and  they fitted large residuals 
at 6 -- 10 keV  by a broad (FWHM$\sim$ 2 keV) 
Gaussian emission line or, alternatively, by two absorption edges associated with low ionized iron and Fe XXV/XXVI.
%in the like spectral model (a blackbody spectrum plus a Comptonized component) 
%allowed to detect  iron absorption edges associated 
%with Fe \ and Fe XXV a
%at  $\sim$7.1 keV and $\sim$9 keV respectively.   
%large residuals at 6 -- 10 keV 
 However, in the framework of this model,  \cite{dai2006} found no evidence of broad, or narrow Fe K lines, between  6 and 7 keV.
%Accounting low energy resolution of RXTE we have not an opportunity for testing any absorption edges. 
But using our model {\it wabs*(blackbody+COMPTB+Gaussian)} we found  an iron line feature during all {\it Beppo}SAX and {\it RXTE} observations.
% as well as obtained the improvement  of fits with additional $Gaussian$ line component at 6 -- 7
%keV in frame of our spectral model.

On the $top$ of Figure~\ref{BeppoSAX_spectra} we demonstrate  the best-fit {\it Beppo}SAX
spectrum  using  our   model  and  in the %left {\it bottom panel} : the best-fit spectrum and 
% $\Delta \chi$ for the model fit without the line component (reduced $\chi^2$=2.15 for 445 d.o.f) and on 
{\it bottom right} panel we show the best-fit spectrum along with 
 $\Delta \chi$ for the model
 % fit with  an addition of  {\it Gaussian}  (K$_{\alpha}$-line) component and  $Bbody$ component at low energies 
 (reduced $\chi^2$=1.16 for 445 d.o.f). In particularly we  find that  an addition of  the soft blackbody-like  
component  of  temperature $T_{BB}=$0.5$-$0.7 keV to the model  significantly improves  the fit quality of the  
{\it Beppo}SAX  spectra.
 % and we show  this effect  in the bottom panels of  Fig.~\ref{BeppoSAX_spectra}.  
%More specifically, %Thus (in this way),
%In particular
%  $COMPTB$ spectral component has the following parameters:
%temperature of the seed photons $T_s$,
%energy index of the Comptonization spectrum $\alpha$ ($=\Gamma-1$), 
%electron temperature $T_e$,   Comptonization  fraction $f$ [$f=A/(1+A)$]
%and the normalization of the seed photon spectrum $N_{COMPTB}$. 
%{\it 
%As discussed above,
%}
 The   line emission is clearly  detected in the range from  5 to  8 keV
 as one can see from the left bottom panel  of   Fig.  ~\ref{BeppoSAX_spectra}. 
 We show  that this line  is  quite broad and it is much wider than 
the instrumental response whose width is smaller than 0.02 keV
\footnote{See ftp://heasarc.gsfc.nasa.gov/sax/cal/responses/98\_11}.
 %({\bf Lenochka, please estimate the lower limit  of the width of  iron line using the instrumental response}).  
%{\it 
%Thus,
%}
Thus  we   include a simple {\it Gaussian} component  
 whose  parameters are  a centroid line energy $E_{line}$, the width of the line $\sigma_{line}$  
and the normalization $N_{line}$  in the model to fit the data in the 6 -- 8 keV  range.  
%(see Fig. \ref{BeppoSAX_spectra}). 
We also include  the interstellar absorption with a column density $N_H$ in the model.
It should be mentioned  that we  fixed certain parameters of the $COMPTB$ component: 
$\gamma=3$ (low energy index of the seed photon spectrum) and $\delta=0$ because we neglect an efficiency  of the  bulk inflow effect vs the  thermal Comptonization   in the case of  NS source 4U~1728-34.

 % applied to outgoing spectrum. 

%An improving the deviations for {\it Beppo}SAX spectral fit at soft energies 
%allows thus to detect an additional spectral component as {\it blackbody} with the temperature 
%$T_{BB}=$~0.5 -- 0.7 keV. 

%It results in the 
%The broad band source spectra were modeled in XSPEC with 
%additive model consisting of following components: i) a $Black body$ at low energies
%(White et al. 1988) described by a temperature $T_{Bbody}$
%and normalization $N_{Bbody}$; ii) a Comptonized component $CompTB$ (Farinelli et al., 2008) 
%for the hard energy part described by
%temperature of the seed photons $T_s$,
%energy index of the Comptonization spectrum $\alpha$ ($\alpha=\Gamma-1$), 
%electron temperature $T_e$,
%index of the seed photon spectrum $\gamma$, 
%bulk parameter $\delta$,  
%logarithm of the illuminating factor 
%parameter $log(A)$, related to the Comptonized fraction $f$ [$f=A/(1+A)$], and 
%normalization of the seed photon spectrum $N_{CompTB}$ and
%iii) a $Gaussian$ component with centroid line energy $E_{line}$, 
%the width of line $\sigma_{line}$ and the nolmalization $N_{line}$ for describing line feature at 6 -- 7 keV energy.  
%The interstellar absorption at low energies with a column density $N_H$ was applied to outgoing spectrum. 

%The  best-fit parameters of  4U~1728-34 spectrum with {\it Beppo}SAX are 
%presented in Table 3. 
For the {\it Beppo}SAX data 
%observations 
(see Tables 1, 3) we find that the spectral index $\alpha$ 
is %around 1 
of 1.03$\pm$0.04
(or the corresponding photon index $\Gamma=\alpha+1$ is %about 2
 2.03$\pm$0.04). 
% in frame of adopted spectral model. 
While  the  temperature of the seed photons $T_s$  of the $COMPTB$ component changes  from 1.2 to 1.3 keV and  
color  temperature of the  
soft  {\it Blackbody} component $T_{BB}$ is around 0.6 keV. 

Unfortunately  {\it RXTE} detectors cannot provide well calibrated spectra  below 3 keV  while the  broad energy band of {\it Beppo}SAX telescopes allows us to determine  the parameters 
of {\it blackbody} components  at soft energies.  
Thus, in order to fit the {\it RXTE} data  we have to fix the  temperature of {\it blackbody} component at a value of 
$T_{BB}=$0.7 keV obtained as an upper limit  in  our   analyze of   the {\it Beppo}SAX data.
% from  4U~1728-34. 
The  best-fit spectral parameters   using {\it RXTE}  observations are presented in Table 4. 
In particular,  we find that electron temperature $T_e$ of the  $COMPTB$ component varies from 2.5 to 15 keV,  while photon   index $\Gamma$  is almost constant ($\Gamma=1.99\pm 0.02$ )
%is around 2)
 %of 2 (see Table 4)
 for all observations.  It is worth noting that the width $\sigma_{line}$ of $Gaussian$ component does not  vary much  and it is  in the range of 0.3 -- 0.6 keV.
% during  outburst transitions. 
Color temperature $T_{s}$  of $COMPTB$ component  is around  1.3 keV which is consistent with that using  the {\it Beppo}SAX data set 
of our analysis (Table 3) and previous studies by  \cite{disalvo2000a} and  \cite{piraino00}.

%Note that the formal freezing $T_BB$ an $T_s$ with lower/higher values led to increasing/decreasing of  electron temperature $T_e$ of Comptonized cloud and make worse fit quality. 

% uplimit) for  
%because of poor responce of $RXTE$ detectors below 3 keV.
%for all available {\it RXTE} data  and thus  we fix a value of $kT$ at  1 keV. 
%When the parameter  we fixed $\log(A)=1$  (see Table 3, 4), as the 
We fixed the value of the   $COMPTB$ parameter $\log(A)$  to 1 when the best-fit values of $\log(A)\gg1$  because in any case of $\log(A)\gg 1$  a  Comptonization fraction $f=A/(1+A)$ is  approximately 1  and  variations  of  $A\gg 1$ do not improve fit quality any more. 
We use a value of hydrogen column $N_H=2.73\times 10^{22}$ cm$^{-2}$, which was found by~\cite{piraino00}.  Systematic error of 0.5\% has been applied to all analyzed {\it RXTE}  spectra.

%Advantages of our additive spectral model with tuned by $Beppo$SAX hints (results) %set forthed 
%parameters applied to RXTE data  (Table 2 and 4) 
%are demonstrated in Figures~\ref{rxte_hard_state_spectrum}, \ref{rxte_soft_state_spectrum} %in comprarison to 
% with reference to %single
% $COMPTB$ model. 
In  Figure~\ref{rxte_hard_state_spectrum} we show an example of 
the best-fit {\it RXTE} spectrum of 4U~1728-34 for the  low luminosity state in units of 
$E*F(E)$ ($top$) and the spectrum in counts  units ({\it bottom panels}) with 
$\Delta\chi$ for the 30042-03-01-00 observation. 
On the  {\it left} bottom panel we demonstrate a fit of  the model $wabs*COMPTB$
%without modelling the line at $\sim$6.7~keV 
%and the $Bbody$  component at low energies 
($\chi^2_{red}$=2.1 for 61 d.o.f) and on the 
$right$,  the same as the latter one but we  add  an iron $Gaussian$ line and 
the $blackbody$ component  using the model $wabs*(blackbody+COMPTB+Gaussian)$ for which we obtain $\chi^2_{red}$=1.18 for 57 d.o.f. The best-fit model parameters for this observation are 
$\Gamma$=1.99$\pm$0.02, $T_e$=10.4$\pm$0.3 keV and $E_{line}$=6.54$\pm$0.03 keV 
(see more details  in Table 4). Red, violet and blue lines stand for 
$blackbody$, $COMPTB$ and $Gaussian$ components, respectively.
We also apply the same procedure to the  spectrum 
%of 4U~1728-34 
during the high luminosity state and  
in the Figure~\ref{rxte_soft_state_spectrum} we present  the results for the 50023-01-12-00 observation.   On the {\it left} bottom   panel  we show a fit 
of  the model $wabs*COMPTB$
%, without modelling the line at $\sim$6.7 keV 
%and the $Bbody$  component at low energies 
($\chi^2_{red}$=1.79 for 61 dof) and 
on the $right$     we present  the same as the latter one but adding  an iron $Gaussian$ line and  
the $blackbody$ components  using the model $wabs*(blackbody+COMPTB+Gaussian)$ 
for which $\chi^2_{red}$=1.2 for 57 d.o.f. The best-fit model parameters in this case are 
$\Gamma$=1.99$\pm$0.02, $T_e$=5.5$\pm$0.1 keV and $E_{line}$=6.75$\pm$0.04 keV 
(see more details in Table 4). 

The adopted spectral model shows a very good performance throughout
all data sets used in our analysis. Namely, a value of reduced
$\chi^2_{red}=\chi^2/N_{dof}$, where $N_{dof}$ is a number of degree of freedom, 
is  less or around 1.0 for most observations. For a small 
fraction (less than 3\%) of spectra with high counting statistics
$\chi^2_{red}$ reaches 1.5. However, it never exceeds a rejection 
limit of 1.7. 
%%%%BEGIN
%{\it
Note that the energy range for the cases, in which we obtain  the  poor fit statistic (two among 127 spectra with $\chi^2$=1.7 for 44 dof),   are related to iron line region.  
%Possibly, it is caused with complexity of iron line shape and relatively poor 
%energy resolution of RXTE. 
It is possibly that shape of Fe line is more complex than a simple Gaussian (i.e. blend of different 
energies, presence of an edge, or broadening by Comptonization). The fits tend to favor a broad line  (see Table 4), which might be caused by Comptonization. However, this possible complexity is not well  constrained by our data. 

Moreover, recent analysis of high-resolution XMM-Newton spectra of 
4U~1728-34 [\cite{ng10}, \cite{egron11}] using  different spectral models also reveal 
evident residuals at 6 -- 7 keV, which are attributed  to the 
presence of a broad iron emission line. This feature can be equally well fitted by 
 a composition of  pure iron line  and corresponding absorption edge  as well as $Laor$ instead of $Gaussian$  line profile [\cite{ng10}] and also by  a relativistically 
smeared line component   or by a  relativistically smeared reflection model component  [\cite{egron11}]. 
This variety  of the line  models using in the data  analysis for  4U~1728-34
%Under the hypothesis 
%that the iron line is produced by reflection from the inner accretion disk, thwe can infer important 
%information on the physical parameters of the system, such as the inner disk radius, Rin = 25?100 km, 
%and the inclination of the system, 44ø < i < 60ø.
%, considering plural line (and corresponding 
%absorption edge) composition as well as $Laor$ instead of $Gaussian$ line profile, 
demonstrates complexity of the line appearance in this source.
%and 
%ambiguousness (
%uncertainty of the  line model for  4U~1728-34.
%}
%%%%END

It is worth noting that we find some differences 
between our values of the best-fit model parameters and those in the literature.
%  for the same set of the  $Beppo$SAX observations. 
  In particular, the photon index  $\Gamma$, estimated by \cite{disalvo2000a} 
  %Di Salvo et al. (2000)
   for observation  id= 20674001, is 1.60$\pm$0.25 while our value of $\Gamma=1.9\pm0.2$. 
This  difference of the index values  can be explained  using slightly  different models. 
\cite{disalvo2000a} included
% Di Salvo et al. (2000) 
 a narrow Gaussian line around 1.7 keV (radiative recombination emission from Mg~XI) 
 in order  to fit an excess of the residuals  of the  continuum  model. 
However our model result,  using  the  $Beppo$SAX 
observation (id=20889003),    confirms the result of  \cite{piraino00}
%Piraino et al. (2000)  
although  we apply a slightly  different spectral  models.  Our  best-fit  photon index 
$\Gamma=1.9\pm0.2$ are very close to that obtained  by  \cite{piraino00} using the best-fit  parameters of COMPTT model [see \cite{tl94}]  electron temperature $kT_e$=3.16$\pm$0.03 keV   and optical depth (for spherical geometry)  $\tau_0=11.4\pm0.2$.
%  found.
 %Piraino et al. (2000). 
%corresponds to the same as our photon index $\Gamma$ 
%value ($\Gamma$=1.9$\pm$0.2) within the limits of error bars.

Thus   using broad band 
{\it Beppo}SAX observations we can well determine all parameters 
of our spectral model 
%the model components of 4U~1728-34 spectrum. 
while  due to  the extensive observations of 4U 1728-34 by {\it RXTE}  we  are able to  
investigate the overall pattern of the source behavior during the spectral transitions 
in the 3 -- 60 keV energy range.  

\section{Evolution of X-ray  spectral properties during spectral state transitions \label{evolution}}

We have established  common characteristics of the  rise-decay spectral transition  of 4U~1728-34 
based on their  spectral parameter evolution of X-ray emission  in  the energy range from 3 to 60 keV  using {\it RXTE}/PCA\&HEXTE data.  
In Figures \ref{rxte_hard_state_spectrum}$-$\ref{rxte_soft_state_spectrum} we present typical examples  
of the {\it RXTE} low and high state  spectra for 4U~1728-34. In fact, one can clearly see  from these Figures % \ref{rxte_soft_state_spectrum}$-$\ref{rxte_hard_state_spectrum}  
that the normalization of thermal (blackbody-like) component is a factor of  2 higher in the high state  than that in the low state, although  photon indices $\Gamma$ for each of these spectra are 
slightly  variable from 1.8 to 2.1 but  mostly concentrated  around $\Gamma$=2 (see that distribution of  $\Gamma$ on the {\it  left-hand} panel of Figure~\ref{hist}). 

We test the hypothesis  of  $\Gamma_{appr} \approx 2 $ using  $\chi^2$-statistic criterium.
% by minimization of corresponding 
%residuals for all measurements. 
We show the distribution of $\chi^2_{red}(\Gamma_{appr})=\frac{1}{N}\sum_{i=1}^N\left(
\frac{\Gamma_i-\Gamma_{appr}}{\Delta\Gamma_i}
\right)^2$  versus of $\Gamma_{appr}$ on the {\it right-hand} panel of  Figure~\ref{hist}.
One can clearly see  a sharp minimum of  function $\chi^2_{red}(\Gamma_{appr})$ around 1 
which takes place in the range of 
%$\Gamma_{appr}=1.99\pm0.02$ for d.o.f=126 with null hypothesis probability $10^{-8}$.
%It is worth noting that 
$\Gamma_{appr}=1.99\pm0.01$ with a confidence level 67\%  and 
$\Gamma_{appr}=1.99\pm0.02$ with a confidence level 99\% for 127 d.o.f. 

% with 
%frequency distribution of $\Gamma$ measurements by fit procedure in frame of adopted model}
%[[almost the 
%same and they are  about  2]]. 
It is important to emphasize that the photon index  $\Gamma$  is also independent of the luminosity of blackbody component of $COMPTB$ $L_{39}/d^2_{10}$ and the plasma temperature of Compton cloud $T_e$
% when both of these parameters change by a factor 5 at least 
(see Figs. \ref{outburst_index_norm}$-$\ref{outburst_index_temperature}).
%Moreover, fit distribution of $\Gamma$ value shown in Fig.~\ref{outburst_index_norm} %7 
 %tends to constancy around $\Gamma\sim$2. Therefore we test mentioned approximation 
%($\Gamma_{appr}=const$) with $\chi^2$-statistic criterium by minimization of corresponding 
%residuals for all measurements. Figure~\ref{hist} ({\it right panel}) demonstrates that minimum of sum $\frac{1}{N}\sum_{i=1}^N\left(
%\frac{\Gamma_i-\Gamma_{appr}}{\Delta\Gamma_i}
%\right)^2$ corresponds to narrow range of $\Gamma_{appr}=1.99\pm0.01$, %by the critical %residual level $\chi^2_{red}=1$.
%which is an agreement with theoretical predictions (see FT11 next section) as well as with previous investigations.

Using {\it Beppo}SAX  data  FT11  suggested  that the  photon index $\Gamma$ is about 2 for quite a few    NS sources which are observed in the different spectral states.  FT11 characterize  the spectral state by a value of electron temperature $T_e$  and they show that $\Gamma=2\pm 0.2$ (or 
$\alpha=1\pm 0.2$)  when  $kT_e$ changes from 2.5 to 25 keV.

%$\sim$
%%Ten days before radio  outburst, 4U~1728-34 being  in IS
% the {\it intermediate} state, 
%%shows  some sign  of  decrease of soft (ASM) X-ray flux (see  Fig.~\ref{spec_evol_R2}) followed by 
%%X-ray  flux rise reaching  its  maximum  just  two days before the radio flare.
% Proper ouburst 
%begins with a soft (ASM, 2-12 keV) X-ray local spike (Fig.~\ref{spec_evol_R2}, R2 set; 
%Fig.~\ref{spec_evol_R6}, R6 set) by some 
%days before radio outburst peak. 
%%Note at the moment of radio peak  the X-ray flux reaches its minimum  (BMC normalization, $N_{BMC}\sim0.7\times10^{-3}$)  when X- spectrum becomes harder  (photon index $\Gamma\sim 1.9$).

A number of X-ray flaring episodes of 4U~1728-34 have been  detected with {\it RXTE} during 1998 -- 1999 
($R3$, $R4$  sets)  and 2000 ($R7$, $R8$ sets) with good rise-decay coverage.  We have searched for  common 
spectral and timing features which can be revealed during these spectral transition episodes.  
We present the combined results of  the spectral analysis of these observations  using 
 our  spectral model  $wabs*(blackbody+COMPTB+Gaussian)$  in
  Figures \ref{lc_1998}$-$\ref{evolution_lc_3}.   
  {\it RXTE}/ASM  count rate is shown on the top panel. 
Further, from the  top to the bottom,  we show the  model flux in two energy bands  3$-$10 keV ({\it blue points})   
and 10 -- 60  keV  ({\it crimson  points}).  In the next panel  we show 
a change of the  transition layer  electron temperature $kT_e$. One can clearly see the 
spectral transition from the high state to the low state   during the time period from MJD 51070 to  
MJD 51215 when electron temperature $kT_e$ varies from 3 keV to 15 keV.  
%On the other hand in the time period from 
%MJD 51195  to  MJD 51210 the source undergoes the transition  from the low state  to the high  state    when $kT_e$ decreases from 15 keV to 4 keV.
 %/seed disk photon/bbody temperature 
%({\it green/crimson/blue}) and
Normalization of  the $COMPTB$ ({\it crimson} points) and   $blackbody$ component 
({\it blue} points)
% [[and $Gaussian$ ({\it green} points)]] 
 are shown in the next panel  of 
%{\it
%next-to-
%}
%bottom panel of 
Figs.~\ref{lc_1998} and \ref{evolution_lc_3}. 
In particular, one can see from Figures \ref{lc_1998}$-$\ref{evolution_lc_3}  
how $COMPTB$ normalization $N_{COMPTB}$ correlates with  
the variations of  ASM count rate and  the model flux in 3-10 keV energy band. 
On the other hand, the normalization of the $blackbody$ component $N_{BB}$ is almost constant 
except at   the flaring episode peak, when $N_{BB}$ increases from 0.09 to 0.27 
(see blue points   in Figure \ref{lc_1998}    at MJD=51093 and 51133).
Moreover  these flare spectral transitions are related to 
a noticeable  increase of soft flux, in the energy range 3-10 keV, and decrease of hard flux, 
that in 10-60 keV (see the second panels from above in Figs. \ref{lc_1998}-\ref{evolution_lc_3}).

The spectral  index $\alpha$ ($\alpha=\Gamma-1$) is presented in the bottom panels of 
 Figures \ref{lc_1998}$-$\ref{evolution_lc_3}. 
 %One can see that  
 The index $\alpha$  slightly varies around 1 (or $\Gamma\sim 2$). 
 Photon index $\Gamma$ variation over the entire set of the observations is presented in  Fig.  \ref{hist}.
The electron temperature  $kT_e$ steadily  decreases during the burst  rise (see {\it blue vertical strips} in Figs. ~\ref{lc_1998} and \ref{evolution_lc_3}).
% all observations demonstrates systematic dropping of $kT_e$ , when 
%the $Blackbody$ flux increases. 
Equivalent width  of iron line $EW_{line}$ and the normalizations of  COMPTB  $N_{COMPTB}$ and   blackbody components  $N_{BB}$  steadily increase  when  electron temperature decreases (see Fig.~\ref{EW_norm_temperature}).

In fact, a decrease of electron temperature $T_e$ of the Compton cloud (transition layer) with an increase of the (disk) soft flux is a well known effect and it is explained in details by 
%    It must be emphasized that this increasing of blackbody-like fluxes is well known 
\cite{tlm98} and \cite{tf04}.
%Titarchuk et al. (1998) and Titarchuk & Fiorito ( 2004))

%  ({\it green/blue/red}) 
%was presented. It is notably that 

%Typically, the electron temperature increases 
%from 4 keV to 12 keV during outburst rise and steeply decreases during outburst peak. 
%The temperature of disk seed photons is also constant on the level of 1.3 keV. 
%As we see in Fig.6 at the vicinity of outburst peak the [3-10] keV and [10-60] keV fluxes shows 
%anticorrelate behaviour pattern.

As shown on the right hand panel of  Figure \ref{outburst_index_norm} 
%\ref{outburst_index_temperature} and 
 %along with quasi-constancy of $Gamma$ vs the electron temperature $kT_e$ and $COMPTB$-normalization 
%$N_{COMPTB}$,  
the Comptonization  fraction $f$  varies from 0.6 to 0.9. 
%Consequently, 
%Comptonization  fraction $f$ shown in the right hand  panel of Fig.~\ref{outburst_index_norm} is 
 %high in most of cases. 
 This means  that  in most  cases the soft disk  radiation  of 4U~1728-34 
 is subjected to reprocessing in Compton cloud  and only a small fraction of 
disk emission component ($1-f$) is directly seen by the Earth observer. 
Thus the energy spectrum of 4U~1728-34 during all states is dominated by a Comptonized component %seen  as a  power-law hard emission in the energy range from 3 to 60 keV, 
while  the direct disk emission is %not seen 
always weaker and detectable in the flaring episodes  only  
(see Figs.~\ref{lc_1998}$-$\ref{evolution_lc_3}).  
%and down panel of Figure~\ref{EW_norm_temperature}).
                                                             
% Thus this (said, introduced) 
%definition of spectral state for NS case based on the cut-off energy of the spectrum.Insofar as in 4U~1728-34  
%the transition  from  the hard (low) to  soft (high) states takes place  
%when electron (plasma)  temperature changes  from  15 keV to 2.5 keV,   
%$-$\ref{evolution_lc_3}).
%6, 7). 
%following FT11 suggestion, we define  the spectral state 
%transition in a NS source  
%can be considered (defined, sutable to consider) 
%in terms of the electron temperature $T_e$ of the Compton cloud (TL). In this case 
%the  hard (low) state is characterized by high electron temperature $T_e$, while 
%the soft (high) state is related to  low $T_e$.  
%electron temperature $T_e$. 
%Note that the electron temperature $T_e$ is a directly measurable quantity and 
%it corresponds to  cut-off energy of the spectrum.

Note that for BHs a definition of spectral transition is related to a change of photon index $\Gamma$ (see e.g. ST09).  However there is no one-to-one correspondence between $\Gamma$  and cutoff  (or efold) energy $E_{fold}$. \cite{ts10} demonstrate using {\it RXTE} data for BH binary XTE J1550-564 that   $E_{fold}$ decreases when $\Gamma$ increases from 1.4 to $2.1-2.2$ until    $\Gamma$ reaches 2.2   and then   $E_{fold}$ increases.  
%There spectral hardness during burst spectral transition. 
Thus {\it for a BH the main parameter used for the spectral transition definition    
is a {\it variable} photon index $\Gamma$ which monotonically increases when a  BH source goes to the high state}.

It is important to emphasize once again that in the NS binary 4U~1728-34   the transition  from  the low state to the  high state  takes place  
when electron (plasma)  temperature changes  from  15 keV to 2.5 keV,   
%$-$\ref{evolution_lc_3}).
%6, 7). 
thus following FT11 suggestion, we define  the spectral state 
transition in a NS source  
%can be considered (defined, sutable to consider) 
in terms of the electron temperature $T_e$ of the Compton cloud (TL). In this case 
the  low state is characterized by high electron temperature $T_e$, while 
the high state is related to  low $T_e$.  
%electron temperature $T_e$. 
Note that the electron temperature $T_e$ is a directly measurable quantity and 
it corresponds to  cut-off energy of the spectrum.

%Using   Figure \ref{outburst_05_rise}  one  can see a  long term quasi-periodicity of the mean flux 
%in  4U~1728-34 with a period 
%(interval, cycle) 
%around six years.   

%{\it
%We have  made an analysis of {\it RXTE} observations  of 4U~1728-34  during four years 
%for 10 intervals  which is indicated with blue rectangulars of Figure~\ref{outburst_05_rise}.  
%and   \ref{evolution_lc_3}).
%}

%The COMPTT model fits the observable spectra well. 
%However, some care should be exercised with the physical interpretation of the 
%best-fit COMPTT parameters, namely the electron temperature.
%  strictly speaking, can be applied for the thermal Comptonization
% (LHS) only. 
% This is not by chance because the 
%In LT99 it was shown that Generic Comptonization spectra (GCS) can be formed 
%as  a result of the combined thermal and bulk Comptonization effects 
%and that they are described by the power-law-exponential-cutoff profile as that of COMPTT.  
%The difference between COMPTT and GCS is in the meaning of the exponential cutoff energy which 
%is determined by electron energy $kT_e$ in the COMPTT case  and by the plasma energy (thermal, 
%bulk and maybe nonthermal) in the GCS case. Therefore, the COMPTT model is applicable to
%GCS shape with the caveat that the meaning of $kT_e$ as the electron temperature
% is generalized to include possible effects of bulk-motion Comptonization and a  non-thermal 
%electron energy distribution. However, the power-law spectral index 
%$\alpha~(=\Gamma-1)$ in both cases is a measure of Comptonization efficiency regardless of the exact

%It should be noted that 
Not all NSs show flares.  Only  a few NS binaries 
(such as Z and $atoll-$sources) 
 display   spectral  transitions during the bursts. 
  Atoll-sources, such as 4U~1728-34, usually show 
the flare transitions.  One can   establish   a substantial 
difference between a NSs  and a BHs due to these flare episodes when a  source evolves from the low to the  high state and when plasma temperature of Comptonized region remarkably changes
(like in 4U 1728-34 from 2.5 keV to 15 keV). 
%Specifically,  NSs and BHs show considerably different variations of spectral characteristics.  
Thus  the NS LMXB 4U~1728-34 
shows  a  steady decrease of electron temperature $T_e$ from the low state to the high state 
while the  photon index $\Gamma$ stays around  2. In contrast, in  BH sources we see a steady monotonic growth of  $\Gamma$ which follows by  its saturation (see e.g. ST09).
 %of $\Gamma$.
 % When compact object binary begin transit into 
%outburst it is located at power dominating state with low luminosity (LHS). This state is 
%characterized with energy spectrum dominated by a Comptonized component combined with a 
%weak thermal component. Then, during developing of outburst, we have different situations for 
%BH and NS. In the case of BH transient, it demonstrate a change  of 
%spectral state with growth of its luminosity, i.e. BH particularly goes into thermal 
%dominating state (HSS). Whereas, in the case of NS transient, it does not change the spectral 
%state with growth of its luminosity, i.e. NS is still at power dominating state (LHS, IS). 
%Thus, the type of spectral behaviour pattern of fransient, which is directly measured from observation, 
%is straight characterized the type of compact object, NS or BH.

\section{Correlations between spectral and timing properties during spectral 
 state transitions \label{transitions}}

The {\it RXTE} light curves have been analyzed using the {\it powspec} task from
FTOOLS 5.1. The timing analysis {\it RXTE}/PCA data was performed in  13 -- 30 keV energy range 
using  the {\it event} mode.  The time resolution for this mode is 1.2$\times 10^{-4}$ s. We
generated power density spectra (PDS) in  0.1 -- 500 Hz frequency range
with 0.001-second time resolution. We subtracted the contribution due to
Poissonian statistics. To model PDS we used QPD/PLT plotting package.

%{\it
Previously, timing analysis of 4U~1728-34 was carried out by \cite{disalvo2001}
as a function of source position in color-color diagram for RXTE data ($R1$, $R2$ sets in
our terms). In island part of the color-color diagram (corresponding to the hardest energy
spectra) the power spectrum of 4U~1728-34 shows several features such as a band-limited
noise component presented up to a few tens of Hz, a low frequency quasi-periodic oscillation
(LFQPO) at frequences between 20 and 40 Hz, a peaked noise component around 100 Hz
and one or two QPOs at kHz frequences. During burst evolution (moving along color-color
diagram) in the lower banana (corresponding to the softer energy spectra) they find a very
low frequency noise (VLFN) component below $\sim$1 Hz. In upper banana (corresponding to
the softest energy spectra) the power spectra are dominated by the VLFN with a peaked noise
component around 20 Hz.

We find  a similar  timing behavior of 4U~1728-34 for our data set along  with the energy
spectra. In particular, 
in Figure \ref{ev_PDS_SP}, we show the details of a typical evolution of
X-ray timing and spectral characteristics during X-ray flares. 
The evolution of {\it RXTE}/ASM count rate during the 1998 ($R3$) outburst transition is presented on the top. 
Red/blue points A, B, and C mark moments at MJD = 51122/51128, 51133.27/51133.34 and 51196/51193 before, during, 
and after X-ray outburst  respectively.  
In the lower panels  ($left$ column)  we show  PDSs for 13-30 keV energy band  along with energy spectral diagram 
$E*F(E)$ ($right$ column) for A (top), B (middle) and C (bottom) points  of X-ray light curve.  
The strong noise component with a break at 1 -- 3 Hz and broad QPOs centered in the range  7 -- 10 Hz are seen before 
%%%BEGIN
%{\it 
%(FWHM=11.7$\pm$4.5, rms=6.3$\pm$1.0\%, $\chi^2=$131 for 102 d.o.f with null hypothesis probability of $10^{-8}$) 
%}
%%%END
and after burst 
%%%BEGIN
%{\it 
%(FWHM=6.0$\pm$2.1, rms=10.6$\pm$0.4\%, $\chi^2=$139 for 102 d.o.f with null hypothesis probability of 
%$4\times 10^{-7}$;  parameter errors correspond to 1$\sigma$ confidence level; see panels A1, C1) 
%}
%%%END
%(see panels A1, C1)
but %none of these
the latter feature is not  seen  at  B moment (see panel B1), i.e. at the 
X-ray flare peak. 
%%%BEGIN
%{\it 
During  B1-burst event  one can see   a noise component with the power peak  shifted to higher frequency 
with  respect to that at  A1/C1 events.  In other words, the burst power spectrum, in this case, consists  of  the ``high frequency'' white-red noise 
component with break at $\sim$40 Hz.
%}
%%%END
%Note also that sometimes before and after burst an additional QPO centered near the break of noise component 
%around $\nu_{l}\sim$20 Hz (e.q., see Fig. 9) is
%presented. 
On the right hand side  we present the $E*F(E)$ spectral diagrams (panels A2, B2, C2) related to the 
corresponding power spectra 
(panels A1, B1, C1).  The data are shown by red points and  
 the spectral model components are displayed  by blue, black, and dashed purple lines for $COMPTB$,  
$blackbody$ and $Gaussian$ components respectively.

Specifically, before the burst (A1 red; 30042-03-08-00 RXTE observation, MJD=51122) one can see a broken power law noise 
component with a break at 1 Hz, broad QPO at 20 Hz (described by Lorentzian with %FWHM of 15 Hz
%%%BEGIN
%{\it 
FWHM=11.7$\pm$4.5 Hz, rms=6.3$\pm$1.0\%, $\chi^2=$131 for 102 d.o.f for 67\%confidence level).
%null hypothesis probability of $10^{-8}$) 
%}
%%%END
Later, just before 
the burst (A1 blue; 
30042-03-11-00, MJD=51128), break frequency of the broken power law noise component shifts from 1 Hz to 3 Hz 
 and QPO becomes  less evident, but still visible at 10$-$20 Hz range.  The low-frequency part   of 
$\nu\times power$ diagram %arises  
below 1 Hz increases right  before the burst (A1, blue). 
During the burst (B1 red, 30042-03-14-02, MJD=51133.27; B1 blue, 30042-03-14-01, MJD=51133.34) one can see 
  a  white-red noise component with the break frequency shifted to higher frequency at  about 40 Hz.  
  %with break at $\sim$10 Hz and  
   QPO component is not seen in the power spectrum during the burst at frequencies  80 Hz and below. 
%%%BEGIN 

We should point out once again that  the similar   behavior of 4U~1728-34 was detected previously by~\cite{disalvo2001}  during the 1996$-$1997 transition from $island$ to $banana$ states in color-color diagram which we  call a  ``burst'' transition  in this Paper. Note that according to \cite{disalvo2001}  the power spectrum at the upper  $banana$ state (at the  maximum of the  burst) consists of two noise components, namely VLFN (very low frequency noise) and HFN (high frequency noise). 
In addition to ~\cite{disalvo2001} we have detected a particular burst when the power spectrum is presented only  with a single HFN component (see B1 panel of Figure \ref{ev_PDS_SP}).
%}
%%%END
%We can suggest  that 
%in this case 
%an outflow (wind)  completely  smears out  the timing signal during the burst. 

After the outburst (C1 blue, 30042-03-18-00, MJD=51193; C1 red, 30042-03-20-00, MJD=51196) one can see the same   
features  in $\nu\times power$ plot as that  before the  burst, but with  slightly  different parameters:  
$\nu_{br}\sim$1,  2 Hz  and 
$\nu_{QPO}\sim$ 6,  10 Hz (described by Lorenzian with 
%%%BEGIN
{\it 
FWHM=6.0$\pm$2.1/15.0$\pm$2.9 Hz, rms=8.1$\pm$0.4/10.6$\pm$1.4\%, $\chi^2=$139/143 for 102 d.o.f; }
all parameter errors correspond to 1$\sigma$ confidence level)%; see panels A1, C1) 

%%%END
%,  and FWHM $\sim$ 5, 15 Hz 
%for the blue and red histograms respectively.

In Figure \ref{PDS} we present  the   $\nu\times power$ plot   observed on March 7, 2000  
(50023-01-01-00 {\it RXTE} observation, MJD=51610) during the quite state in order to  compare it with the  
typical sample of  PDS during the X-ray flare event (see  panel B1 in Fig.~\ref{ev_PDS_SP}).
%\ref{evolution_lc_3}).
%Here data are denoted by red points.
%The spectral model, presented in Figure \ref{ev_PDS_SP} by its  components is shown
% by blue, black, and dashed purple lines for the $COMPTB$, $blackbody$ and $Gaussian$ components respectively.

%Note that  the normalization $N_{BB}$ of the $blackbody$ component undergoes noticeable changes during X-ray flare, 
%is  increases from 0.01 to 0.03 
%see also bottom panel of Fig. \ref{lc_1998}),  
%\ref{evolution_lc_3} 
%where one can clearly see that $N_{BB}$ increases from 0.01 to 0.03. 
%Specifically, at the flare peak (moment B) the
%total flux increases at least by a factor of 1.5 with respect to that before the flare, although the photon index of 
 %the $COMPTB$ component $\Gamma$ is constant around 2 for all moments (A, B and C). 
%This  can be seen   with the constancy (stability) 
%In fact,  the shape of $COMPTB$ ($blue$ line) component is stable during  X-ray flare  (see A2, B2, C2 panels). 

%{\it
%In comparison to similar investigations for BHC GRS~1915+105 (see Fig.16 of TS09) during outburst, 
%the spectral characteristics, particularly 
%photon index $\Gamma$, undergo noticable changes, while timing properties are almost 
%the same to that for NS 4U~1728-34.
%}

\section{Discussion  \label{disc}}
We show the quasi-constancy  of th photon (spectral)  index  in quite a few observations of NS source 4U~1728-34  using {\it Beppo}SAX and {\it RXTE} observations.  
In Figures \ref{outburst_index_norm}$-$\ref{outburst_index_temperature} we present plots of the photon index $\Gamma$ as functions of the our model parameters: electron temperature
$T_e$ (in keV),  the  $COMPTB$ normalization (which equals  the normalization of NS blackbody seed photons)  and the Comptonization fraction $f=A/(1+A)$.  
 %Thanks to numerocity of $RXTE$ observations and wide spectral range and good spectral resolution %of  $BeppoSAX$ detectors, we can refine the spectral model of 4U~1728-34 and established the stability  of the spectral index during burst evolutiom. 
We obtain these results using an application of the first principle spectral model to  extensive  {\it Beppo}SAX and 
{\it RXTE}  observations of   the  NS binary source 4U~1728-34. 
FT11 give an explanation  of this index  stability which was also  revealed in other observations of NS binaries. 
In this Paper, for a completeness of our presentation,   we   review   the main points of the FT11 explanation in terms of the transition layer model (TLM). 

The energy balance in the transition layer (TL) is dictated by Coulomb collisions with protons (gravitational energy release), while inverse Compton 
and free-free emission are the main cooling channels [see a formulation of this problem in the pioneer work by \cite{zs69} 
%Zel'dovich \& Shakura 1969 
and also a similar consideration in \cite{bisn80}].   
In fact, for the characteristic electron temperature (3 keV $\la kT_e\la$ 30 keV)  and density values 
($\la 10^{-5}$ g cm$^{-3}$) of these regions in low mass X-ray binaries, Compton cooling dominates over free-free emission and   
the relation between the energy flux  per unit surface area of the corona  $\qcor$,  the radiation energy  density 
 $\varepsilon(\tau)$  and electron temperature $T_e$  is given by [see also \cite{tlm98}]
\begin{equation}
 \frac{Q_{cor}}{\tau_0} \approx  20.2\varepsilon(\tau)T_e(\tau),
 %{C_{comp}\varepsilon(\tau)T_e},
\label{energy_balance}
\end{equation}
where  $\tau_0$ is Thomson optical depth of
the TL. 
The distribution $\varepsilon(\tau)$ is obtained as a solution of the diffusion equation
\begin{equation}
\frac{d^2\varepsilon}{d\tau^2} =-\frac{3\qtot}{c\tau_0},
\label{diffusion_equation}
\end{equation}
where  $\qtot=\qcor + \qd$ is the sum of the corona (TL) and intercepted disk fluxes, respectively. 
 Combination of Eq. (\ref{diffusion_equation}) with two boundary conditions at NS surface and outer TL boundary  leads us  to the formulation of the TL boundary problem (see details in  FT11).
 The solution for $\varepsilon(\tau)$ is then given by
\begin{equation}
 \varepsilon(\tau)=\frac{2\qtot}{c} \left[1+ \frac{3}{2}\tau_0\left(\frac{\tau}{\tau_0} -
  \frac{\tau^2}{2\tau_0^2}\right)\right].
\label{ene_vs_tau}
\end{equation}
In order to establish the average plasma temperature $T_e$  one should estimate the mean energy density in 
the TL as 
%We consider now an average value of the energy density as
\begin{equation}
<\varepsilon(\tau)>=\frac{1}{\tau_0}\int^{\tau_0}_0 \varepsilon(\tau) d\tau=\frac{
\qtot}{c}(2+\tau_0).
\label{average_ene}
\end{equation}
If we now substitute the result of equation (\ref{average_ene}) into equation (\ref{energy_balance}), after a bit of 
straightforward algebra we obtain
\begin{equation}
\frac{\kte \tau_0 (2+\tau_0)}{m_e c^2}=\frac{0.25}{1+\qd/\qcor}.
\label{ktetau}
\end{equation}

One should use a  formula  for  spectral index $\alpha$ 
\begin{equation}
\alpha=-\frac{3}{2}+\sqrt{\frac{9}{4}+\frac{\beta}{\Theta}}, 
\label{alpha_general}
\end{equation}
where $\Theta \equiv \kte/m_e c^2$ and  $\beta$-parameter
defined in \cite{tl95}.
% (see also formula \ref{} here). 
%as it was assumed by the authors. 
If we replace $\beta$ by  its diffusion limit  $\beta_{\rm diff}$ 
\begin{equation}
\beta_{\rm diff}=\frac{1}{\tau_0 (2+\tau_0)}
\label{beta_diff}
\end{equation} 
and using 
%(Eq. \ref{beta_diff}), and  
equation (\ref{ktetau}), we obtain  the diffusion spectral index as 
\begin{equation}
\alpha_{\rm diff}= -\frac{3}{2}+\sqrt{\frac{9}{4}+ \frac{1+\qd/\qcor}{0.25}},
\label{alpha_diff}
\end{equation}
or $\alpha_{\rm diff}\approx1+0.8~ \qd/\qcor$  and 
\begin{equation} 
\Gamma_{diff}\approx1+\alpha_{\rm diff}=2+0.8~\qd/\qcor
\label{gamma_diff}
\end{equation}
 for $\qd/\qcor<1$.
 
 Thus until $\qd/\qcor\ll1$ photon index $\Gamma\approx2$. This is precisely that we see in the observations of NS 4U 1728-34 
 (see Figs. \ref{hist}$-$\ref{outburst_index_temperature}).   
   
However in  BHs  we observe that the photon  index monotonically increases with QPO frequency and mass accretion 
rate and finally saturates (see ST09, TS09 and ST10).  Recently \cite{lt10}, hereafter LT11,  have made  Monte 
Carlo simulations of X-ray spectral formation  in  Compton cloud which surrounds a BH and they reproduce the  
observed  correlation index vs mass accretion rate. 
They also  demonstrate that the index saturation observed in BH sources is a result of two effects, namely   
cooling of the converging flow by the soft disk photons  along with  the photon capture by a BH. 
In fact, spectral  index is the inverse of the Comptonization parameter $Y$  which is proportional to mean 
number of up-scattering $N_{sc}$ and efficiency of up-scattering $\eta$.   But in the relatively cold converging 
flow (CF) photons  (when mass accretion rate $\dot m$ in Eddington units is much greater than 1) mostly up-scatter 
off electrons  in the direction of the flow for which   $N_{sc}$ is proportional to CF optical depth $\tau_{CF}$ 
(or $\dot m$) and $\eta$ is inverse proportional to   $\tau_{CF}$ ($\dot m$).  Thus the spectral (photon) index saturates when mass accretion rate increases.  This is precisely what was reproduced in the MC simulations by LT11. {\it  Hence one can conclude that the monotonic growth of the photon index
  $\Gamma$ with mass accretion rate followed by its saturation is the observational signature of a black hole while the constancy of  $\Gamma$ (around 2) versus $\dot m$ (or electron temperature) is the NS signature.}  
 
Recently \cite{soria10} found  that in ultra luminous  X-ray source HLX1 photon index changes from 1.8 to 2.95 but they can not find an argument either this source  is intermediate-mass BH or foreground NS.    Comparison of our  and FT11 results for NSs and ST09 results for BHs 
(see also LT11)  probably  indicates that HLX1 is a black hole because its photon index changes in the wide range  from 1.8 to 2.95  (see Fig.  \ref{bh_ns_examples})  but  in  a NS case the index does not vary and has  almost constant value  around 2  (see Figs. \ref{hist}$-$\ref{outburst_index_temperature}).
%  and monotonic growth with 
%normalization when the source transits from hard (low) to soft (high) states, In contrast to BHs, NSs, in particular 4U~1728-34 
%during outburst,  typically show the stability of spectral index mainly at 1.

%For example recently, Soria et al. (2010, astro-ph/1008.3382v1) discuss ...

\section{Summary \label{summary}} 

We presented our analysis  of the spectral properties observed in X-rays from 
the 
%{\it
%burstimg atoll source
%}
neutron star  X-ray binary  4U~1728-34 during  transitions
between the low state  and the high state. We analyze a number of  transition episodes 
from this source  observed with {\it BeppoSAX}  and 
{\it RXTE} satellites.  
For our  analysis we use   a good spectral coverage and  resolution of  $BeppoSAX$ detectors 
from 0.1 to 200 keV along  with $extensive$  $RXTE$ observations in the energy range from 3 to 60 keV.    

We show that the X-ray broad-band energy spectra during all spectral states can be adequately fitted by   composition of  the {\it Blackbody}, Comptonized component ({\it COMPTB}) and 
 {\it Gaussian} component. 
 We  also show  that photon index $\Gamma$ of the best-fit Comptonized component  in 4U~1728-34 is 
almost constant, about 2 (see Fig. \ref{hist}) and consequently almost 
 independent of   {\it COMPTB} normalization $L_{39}/D^2_{10}$  which is proportional to the disk mass accretion 
rate $\dot m$,  (see the left hand panel of Fig. \ref{outburst_index_norm}) and plasma temperature of 
Compton cloud $T_e$  (see Fig. \ref{outburst_index_temperature}).  Note the soft (disk) photon luminosity 
$L_{39}$ is units $10^{39}$ erg s$^{-1}$ and distance to the source $D_{10}$ is units of $10$ kpc. 
 We  should remind a reader  that this index stability  has been recently suggested using  a quite 
a few number of other  NS sources, Cyg X-2, Sco X-1, GX 17+2,
GX 340+0, GX 3+1, GX 349+2, X 1658-298, GS 1826-238, 1E 1724-3045 which were been observed by {\it Beppo}SAX 
at different spectral  states  [see details in  FT11].
 
 %We also present an observable 
 %[independence]
%{\it
%quasi-constancy
%}
 % of  photon index $\Gamma$ when  the (disk)  mass accretion rate $\dot m$ or  the {\it COMPTB}  normalization, which is proportional to $\dot m$, varies  by  a factor 5 
 % (see  Fig.~\ref{outburst_index_norm}).
 
A relatively  high value of  Comptonized   fraction $f=0.6-0.9$, obtained in the framework of our spectral model,   indicates 
%gives us  a strong  argument  
to  significant reprocessing  of X-ray disk emission in Compton cloud in  4U~1728-34.    
%  The use of th ({\it COMPTB}) normalization, which is 

We also find using {\it Beppo}SAX  observations  that there are two sources of blackbody emission, 
one is presumably related to   the accretion disk  and another one is related  to NS surface for which  
temperatures of soft photons are about 0.7 keV and 1.3 keV, respectively.   

We demonstrate  that
 the photon index $\Gamma\sim 2$ is almost constant  when  the source moves from the low state to the high state, in other words  when the plasma temperature of Comptonized region varies from 15 to 2.5 keV  (see  Fig.~\ref{outburst_index_temperature}). 
% the transition to the soft state. 

%direct BB spectrum $kT_{BB}$ is always less than the temperature of the thermally Comptonized 
%seed photons $kT_s$ for 4U~1728-34.
%which  is also  in agreement with power density spectra seen with a power law component.

% We  also show  that photon index $\Gamma$ of the best-fit Comptonized component  in 4U~1728-34 is 
%almost constant, about 2 (see Fig. \ref{hist}) and consequently almost 
% independent of   {\it COMPTB} normalization $L_{39}/D^2_{10}$  which is proportional to the disk mass accretion 
%rate $\dot m$,  (see the left hand panel of Fig. \ref{outburst_index_norm}) and plasma temperature of 
%Compton cloud $T_e$  (see Fig. \ref{outburst_index_temperature}).  Note the soft (disk) photon luminosity  $L_{39}$ is units $10^{39}$ erg s$^{-1}$ and distance to the source $D_{10}$ is units of $10$ kpc.  We  should remind a reader  that this index stability  has been recently suggested using  a quite 
%a few number of other  NS sources, Cyg X-2, Sco X-1, GX 17+2,
%GX 340+0, GX 3+1, GX 349+2, X 1658-298, GS 1826-238, 1E 1724-3045 which were been observed by {\it Beppo}SAX 
%at different spectral  states  [see details in  FT11].

We present the strong theoretical arguments  that the dominance of the energy release in the transition layer (TL) 
%  thermal Comptonized component formed in the 
%transition layers 
 with respect to the soft  flux coming from the accretion disk,  $Q_{disk}/Q_{cor}\ll1$  leads
to almost   constant photon  index $\Gamma\approx2$.

 Thus we argue that {\it this index stability  is the intrinsic signature of a NS binary source while in BHs the index monotonically changes with mass accretion rate and ultimately saturates} (see ST09).
 In Figure \ref{bh_ns_examples} we demonstrate  the index correlation vs mass accretion rate for a number of  BH sources and how  the index depends on mass accretion rate in NS 4U 1728-34. 
Photon indices  of BH candidate sources  (GRS~1915+105, GX~339-4, SS~433 and GRO~J1655-40)   show  clear correlation with mass accretion rate $\dot m$ or with soft photon normalization $L_{39}/D^2_{10}$ which is proportional to $\dot m$ .   
This correlation  is followed by the index saturation when  $\dot m$ exceeds a certain level.  
%in comparison to $atoll$ NS source (4U~1728-34) sample 
 The behavior of the index for the  considered NS 4U~1728-34  is clear different 
  from that for the sample of  BHC  sources.   

We acknowledge Chris Shrader and Cristiano Guidorzi for careful reading and editing the  presented paper.  
We are very grateful to the referee for  his/her valuable comments and corrections 
of the content of this Paper.

\appendix
\section{On the definition  of  the normalization of the  COMPTB and BMC Models}

The {\it COMPTB} and {\it BMC} models describe the outgoing spectrum as a convolution 
of the input ``seed'' blackbody-like spectrum, whose normalizations are $N_{COMPTB}$ and
$N_{BMC}$ and color temperature is $kT$,  with the  Comptonization Green's function. Similarly 
to the ordinary {\it bbody} XSPEC model, bolometrical luminosity

\begin{equation}
L_{bol}=\int_0^{\infty} E\times A(E) dE,
\end{equation}
where 
$A(E)$ is the photon flux density of blackbody radiation 
\begin{equation}
A(E)=8.052 \times K \times \frac{E^2}{(kT)^4}\times(\exp{\frac{E}{kT}}-1)^{-1},
\end{equation}
and 
%kT - color temperature in keV (par1 for both COMPTB and BMC models), and
$K=N_{COMPTB} ~ N_{BMC}$ is the normalization of the seed blackbody photon spectrum, defined in the same way as the XSPEC bbody model.% (par7 for both COMPTB and BMC models).

%$K=N_{COMPTB},N_{BMC}=L_39/D_10^2$ (par2 in both COMPTB and BMC models),
%par7 = Cn, normalization of the seed photon spectrum, defined in the same
%way as the XSPEC $bbody$ model

Thus, one can calculate the emergent luminosity of the source as an integral
\begin{equation}
L_{bol}=8.052\times K\int_0^{\infty}\frac{z^3dz}{e^z-1}=8.052\times K \times \frac{\pi^4}{15}.
\end{equation}

On the other hand $N_{BMC},N_{COMPTB}=K=L_{39}/D_{10}^2$
[section 6.2.10 of ``User's Guide of an X-Ray Spectral Fitting Package XSPEC v.12.6.0''
\footnote{https://heasarc.gsfc.nasa.gov/docs/software/lheasoft/xanadu/xspec/XspecManual.pdf}  
%\footnote{http://heasarc.gsfc.nasa.gov/xanadu/xspec/manual/Additive.html} 
and 
\cite{F08}\footnote{http://heasarc.gsfc.nasa.gov/docs/xanadu/xspec/models/comptb.html}]
%http://heasarc.gsfc.nasa.gov/xanadu/xspec/manual}
%\footnote{http://heasarc.gsfc.nasa.gov/docs/software/lheasoft/xanadu/xspec/models/comptb.html}
%section 6.2.10 of ``User's Guide of an X-Ray Spectral Fitting Package XSPEC for version 12.6.0''
%\footnote{http://heasarc.gsfc.nasa.gov/xanadu/xspec/manual/Additive.html}
, where 
$L_{39}$ is the source luminosity in units of $10^{39}$ ergs$^{-1}$ and  
$D_{10}$ is the distance to the source in units of 10 kpc; 
 and if we know K we can find $L_{39}$ having $D_{10}$. 
Thus, similarly to the ordinary {\it bbody} XSPEC model, the normalizations $N_{COMPTB}$ and $N_{BMC}$ 
are a ratio of the source (disk) luminosity to the square of the distance
\begin{equation}
N_{BMC},N_{COMPTB}=\biggl(\frac{L}{10^{39}\mathrm{erg/s}}\biggr)\biggl(\frac{10\,\mathrm{kpc}}{D}\biggr)^2.
\label{bmc_norm}
\end{equation}

This implies important property of both COMPTB and BMC models.  Namely 
using these models one  can correctly  evaluate  the normalization of the original ``seed''
component, which is presumably a correct mass accretion rate indicator.

\newpage

\begin{deluxetable}{cccc}
%%%%%\rotate
\tablewidth{0in}
\tabletypesize{\scriptsize}
%  \begin{center}
    \tablecaption{The list of $Beppo$SAX observations of 4U~1728-34  used in analysis.}
    \renewcommand{\arraystretch}{1.2}
%    \begin{tabular}[h]
%      \hline
\tablehead{
Obs. ID& Start time (UT)  & End time (UT) &MJD interval}
%Satellite&Obs. ID& Start time (UT)  & End time (UT)}
%%%%%Obs.  &ID           & time (UT)& time (UT)& of state& }
\startdata
20674001& 1998 Aug. 23 19:15:27 & 1998 Aug. 24 09:14:15 &51048.8-51049.4$^1$ \\
20889003& 1999 Aug. 19 02:01:32 & 1999 Aug. 20 04:54:32 &51409.1-51410.2$^2$\\
%20889003& 1999 Aug. 19 02:01:32 & 1999 Aug. 20 04:54:32 &51409.1-51410.2$^1$& \cite{piraino00}\\
      \enddata
%      \hline
%      \end{tabular}
   \label{tab:table}
% \end{center}
Reference
$(1)$ \cite{disalvo2000a}, $(2)$ \cite{piraino00} 
%Piraino et al., (2000)
\end{deluxetable}

\newpage
\begin{deluxetable}{ccclll}
%\rotate
\tablewidth{0in}
\tabletypesize{\scriptsize}
%  \begin{center}
    \tablecaption{The list of groups of {\it RXTE} observation of 4U~1728-34 used in analysis.}
    \renewcommand{\arraystretch}{1.2}
%    \begin{tabular}[h]
%      \hline
\tablehead{Number of set  & Dates, MJD & {\it RXTE} Proposal ID&  Dates UT & Rem. & Ref.}
 \startdata
R1  &    50128-50143 & 10073        & Feb. 15 -- March 1, 1996 &   & 1, 2, 3, 4, 5, 7, 8 \\
R2  &    50710-50728 & 20083 & Sept. 19 -- Oct. 1, 1997 &      &3, 4, 5, 7, 8 \\
R3  &    51086-51196 & 30042 & Sept. 30, 1998 -- Jan. 18, 1999 &   & 4  \\
R4  &    51409-51443 & 40019 & Aug. 19 -- Sept. 22, 1999& BeppoSAX &  8\\
R5  &    51237-51359 & 40027        & Feb. 27 -- Jul. 10, 1999 &  & 6, 8\\
R6  &    51198-51213 & 40033        & Jan. 20 -- Feb. 4, 1999& & 6, 8\\
R7  &    51667-51733 & 50023        & March 7 -- Jul. 8, 2000& & 6, 8\\
R8  &    51652-51657 & 50029 & Apr. 18 -- 23, 2000 & & 6, 8 \\
%R9  &                & 50030 & Jan. 28 -- Nov. 15, 2001 &  & 6 \\
%R10  &               & 60029 & May 27 -- 29, 2001 &  &  \\
%R11  &               & 70028 & March 3 -- 5, 2002 &  &  \\
%R12  &               & 90406 & March 12 00:00:06.9 -- ?????, 2004 &  &  \\
%R13  &               & 91023 & March 12, 2005 -- June 27, 2007 &  & this work \\
      \hline
      \enddata
%      \hline
%      \end{tabular}
    \label{tab:par_bbody}
%  \end{center}
References:
(1) Strohmayer et al. 1996; 
(2) Ford \& van der Klis (1998);
(3) van Straaten et al. (2001); 
(4) Di Salvo et al. (2001); 
(5) Mendez, van der Klis \& Ford (2001); 
(6) Migliari, van der Klis \& Fender (2003); 
(7) Jonker, Mendez \& van der Klis (2000);
%(8) \cite{disalvo2001}
(8) TS05
\end{deluxetable}

\newpage
\bigskip
\begin{deluxetable}{ccccccccccccccc}
%\begin{deluxetable}{ccccccccccccccc}
\rotate
\tablewidth{0in}
\tabletypesize{\scriptsize}
%  \begin{center}
    \tablecaption{Best-fit parameters of spectral analysis of {\it Beppo}SAX 
observations of 4U~1728-34 in 0.3-60~keV energy range$^{\dagger}$.
Parameter errors correspond to 1$\sigma$ confidence level.}
%\vspace{1em}
    \renewcommand{\arraystretch}{1.2}
%    \begin{tabular}[h]
%      \hline
%ID               & day  & & &   $\Gamma-1$          &           &$L_{39}/d^2_{10}$& keV & keV   &  &  keV        &  & & & }
 \tablehead
{Observational & MJD, & $T_{BB}$ & $N_{BB}^{\dagger\dagger}$ &$T_s$ & $\alpha=$  & $T_e,$ & $\log(A)$ & N$_{COMPTB}$ &  E$_{line}$,&   $N_{line}^{\dagger\dagger}$ & EW$_{line}$,&  $\chi^2_{red}$ (d.o.f.)\\
ID             & day  & keV         &                           & keV  &$\Gamma-1$  & keV    &                                  &                                   &   keV       &        & eV    &                    }
 \startdata%   id     MJD      kT_Bbody     N_bb     [kT_s]    alf       T_e       log    norm_COMPTB E_line[Sigma_l] N_line Xi_2(dof)  Flux3-10 Fl10-60
20674001   &        51048.70 & 0.47(3)& 2.65(2) & 1.30(3)& 0.99(7) & 3.76(8) & 0.10(4) & 4.18(3) & 7.4(1) &  0.55(4)& 52(16) & 1.25(457)\\
20889003   &        51409.50 & 0.62(5)& 1.61(1) & 1.21(5)& 1.07(4) & 3.29(4) & 1.06(6) & 3.56(2) & 6.0(1) &  0.43(4)& 51(11) &1.16(445)\\     
      \enddata%     \hline
%      \end{tabular}
    \label{tab:fit_table}
%  \end{center}
$^\dagger$ The spectral model is  $wabs*(blackbody + COMPTB + Gaussian)$,
normalization parameters of $blackbody$ and $COMPTB$ components are in units of 
$L_{37}/d^2_{10}$ $erg/s/kpc^2$, where $L_{37}$ is the source luminosity in units of 10$^{37}$ erg/s, 
$d_{10}$ is the distance to the source in units of 10 kpc 
and $Gaussian$ component is in units of $10^{-2}\times total~~photons$ $cm^{-2}s^{-1}$ in line.
%$^{\dagger\dagger\dagger}$when parameter $\log(A)\gg1$, it is fixed to a value 1.0 (see comments in the text). 
%$\sigma_{line2}$ of Gaussian2 component is fixed to a value 0.01 keV (see comments in the text), 
%$^{\dagger\dagger\dagger\dagger}$spectral fluxes (F$_1$/F$_2$) in the (3 -- 10)/(10 -- 60) keV energy ranges, correspondingly, 
%in units of $\times 10^{-9}$ ergs/s/cm$^2$.
\end{deluxetable}

\newpage
\bigskip
\begin{deluxetable}{lllccccccccccccc}
%\begin{deluxetable}{cccccccccccccc}
\rotate
\tablewidth{0in}
\tabletypesize{\scriptsize}
%  \begin{center}
    \tablecaption{Best-fit parameters of spectral analysis of PCA \& HEXTE/{\it RXTE} 
observations of 4U~1728-34 in 3-60~keV energy range$^{\dagger}$. 
The count rate corresponds to 5 PCA units and corrected for background.
Parameter errors correspond to 1$\sigma$ confidence level.}
%\vspace{1em}
    \renewcommand{\arraystretch}{1.2}
%    \begin{tabular}[h]
%      \hline
%ID               & day  & & &   $\Gamma-1$          &           &$L_{39}/d^2_{10}$& keV & keV   &  &  keV        &  & & & }
 \tablehead
{Observational & Start Time, & Exposure& Rate,  &$\alpha=$   & $T_e,$ & log(A)$^{\dagger\dagger}$ & N$_{COMPTB}^{\dagger\dagger\dagger}$ & $N_{Bbody}^{\dagger\dagger\dagger}$ & E$_{line}$,&   $N_{line}^{\dagger\dagger\dagger}$ & EW$_{line}$,&  $\chi^2_{red}$ (d.o.f.)& F$_1$/F$_2^{\dagger\dagger\dagger\dagger}$ \\
ID             &  MJD        & Time, s& cnt/s  & $\Gamma-1$ & keV    &                           &                                       &                                  &  keV       &                                      &   eV&                      &                                                & }
 \startdata%   id     MJD    alf     T_e       log     norm_COMPTB    N_bb    E_line      N_line   Xi_2(dof)  Flux3-10 Fl10-60

10073-01-01-00  &50128.49 &9632 &1824 &0.98(6) &3.54(3)  &0.8(1)   &3.4(1)   &2.4(2)   &6.6(1)   &  1.2(1) & 100(30) &1.16(64) &2.82/1.43 \\   
10073-01-01-000 &50128.494&12180 &1752 &1.00(3) &3.71(4)  &0.7(1)   &3.4(2)   &2.09(4)  &6.68(8)  &  0.7(1) & 106(20) &1.1(64) &2.69/1.49 \\  
10073-01-02-00  &50129.27 &9488 &2071 &1.00(2) &2.91(2)  &1.00$^{\dagger\dagger}$&4.05(6)   &2.85(5)  &6.63(1)  &  0.57(1)&120(20) &0.98(65) &3.26/1.31 \\
10073-01-02-000 &50129.27 &9840  &1927 &1.11(7) &3.18(7)  &0.94(6)  &3.7(2)   &2.56(2)  &6.64(3)  &  0.73(2) & 110(20) &1.28(64) &3.03/1.34 \\
10073-01-03-000 &50129.66 &11550 &2040 &1.02(3) &3.16(4)  &0.41(3)  &4.53(5)  &3.19(2)  &6.58(2)  &  0.32(1) & 150(20) &1.39(64) &3.21/1.28 \\
10073-01-03-00  &50129.93 &9488 &2071 &1.05(4) &3.17(4)  &0.41(4)  &4.56(7)  &3.28(2)  &6.58(7)  &  0.30(1) & 130(10) &1.15(64) &3.22/1.27 \\
10073-01-04-000 &50131.46 &12540 &1493 &0.99(2) &4.64(3)  &0.2(1)   &3.25(2)  &2.67(2)  &6.53(5)  &  0.38(1) & 90(10) &1.32(64) &2.08/1.46 \\
10073-01-04-00  &50131.79 &9664 &1387 &0.95(2) &5.2(1)   &0.18(5)  &3.12(3)  &2.36(3)  &6.50(3)  &  0.32(1) & 90(10) &1.07(64) &2.08/1.46 \\
10073-01-06-000 &50135.48 &4160  &994  &0.91(7) &8.1(3)   &0.27(2)  &2.3(2)   &1.19(3)  &6.50(2)  &  0.30(2) & 50(8)  &1.4(64) &1.45/1.60 \\
10073-01-06-00  &50135.78 &9600   &1021 &0.90(8) &8.2(2)   &0.30(2)  &2.34(6)  &1.18(2)  &6.51(2)  &  0.31(1) & 40(10) &1.1(64) &1.43/1.71 \\
10073-01-08-000 &50136.88 &14560 &1035 &0.90(9) &8.6(3)   &0.33(1)  &2.32(4)  &1.05(1)  &6.50(3)  &  0.32(1) & 40(10) &1.12(64) &1.50/1.81 \\
10073-01-07-000 &50136.89 &16000  &972  &0.86(3) &8.4(1)   &0.29(2)  &2.25(5)  &1.06(3)  &6.48(5)  &  0.31(1) & 40(10) &0.96(64) &1.41/1.73 \\
10073-01-07-00  &50137.22 &2992 &1058 &0.98(2) &8.5(4)   &0.39(3)  &2.43(6)  &1.05(2)  &6.45(4)  &  0.33(2) & 40(10) &1.17(64) &1.54/1.76 \\
10073-01-08-00  &50138.06 &7888 &1030 &0.89(9) &8.1(2)   &0.31(2)  &2.39(3)  &1.07(1)  &6.60(2)  &  0.32(1) & 40(10) &1.56(64) &1.49/1.75 \\
10073-01-09-000 &50138.86 &16510 &1090 &0.91(8) &8.4(1)   &0.33(1)  &2.51(2)  &1.08(3)  &6.48(3)  &  0.34(1) & 50(10) &1.54(64) &1.58/1.89 \\
10073-01-09-00  &50139.20 &7968 &1095 &0.82(9) &6.9(1)   &0.18(1)  &2.66(3)  &1.64(2)  &6.62(2)  &  0.16(1) & 70(10) &1.05(64) &1.73/1.74 \\
10073-01-10-01  &50143.00 &12530 &1191 &0.99(2) &12.2(3)  &0.84(1)  &2.40(4)  &0.92(3)  &6.61(5)  &  0.44(5) & 50(20) &1.09(64) &1.49/2.46 \\
20083-01-01-00  &50710.24 &9520  &2013 &1.0(2)  &3.01(4)  &0.59(5)  &3.8(3)   &2.75(6)  &6.6(1)   &  0.5(2)  & 120(20) &0.86(57) &4.65/4.3  \\   
20083-01-01-01  &50710.51 &4464 &2043 &0.99(7) &2.96(5)  &0.4(1)   &5.0(3)   &2.01(2)  &6.7(1)   &  1.0(1)  & 160(30) &0.79(57) &3.33/1.2  \\  
20083-01-01-02  &50711.31 &4928 &1998 &1.00(5) &2.89(3)  &0.6(1)   &4.65(4)  &2.6(3)   &6.7(1)   &  0.5(1)  & 110(20) &0.8(57) &3.25/1.17 \\
20083-01-01-020 &50712.24 &11500  &1183 &0.99(7) &3.02(4)  &0.3(1)   &4.3(3)   &2.1(1)   &6.58(8)  &  0.33(8) & 110(10) &1.15(57) &3.03/1.15 \\
20083-01-02-000 &50712.35 &3264 &1908 &1.00(5) &3.06(2)  &0.28(5)  &4.2(1)   &2.3(1)   &6.5(1)   &  0.6(3)  & 130(10) &0.81(57) &3.09/1.14 \\
20083-01-02-01  &50712.65 &12770 &1657 &1.00(8) &3.02(3)  &0.3(1)   &4.6(1)   &2.1(2)   &6.6(1)   &  0.3(1)  & 130(8)  &0.89(57) &3.27/1.20 \\
20083-01-03-00  &50714.98 &3520  &2281 &0.99(9) &2.96(2)  &0.5(2)   &5.45(3)  &2.8(6)   &6.7(1)   &  0.4(1)  & 140(10) &0.72(57) &3.71/1.36 \\
20083-01-03-000 &50715.01 &15650 &2431 &1.12(9) &2.85(2)  &1.00$^{\dagger\dagger}$  &5.69(8)  &2.19(1)  &6.7(1)   &  0.3(1)&130(10) &0.93(58) &3.96/1.44 \\
20083-01-04-00  &50717.51 &16780 &1644 &0.99(4) &3.22(7)  &0.15(5)  &3.5(3)   &2.2(1)   &6.59(9)  &  0.33(9) & 100(10) &0.92(57) &2.65/1.01 \\
20083-01-04-01  &50718.38 &16420 &1388 &0.99(2) &4.21(8)  &0.09(3)  &3.1(1)   &1.85(9)  &6.5(1)   &  0.5(3)  & 90(10) &1.39(57) &2.19/1.10 \\
20083-01-03-02  &50721.18 &6736 &1002 &0.99(3) &3.02(3)  &0.28(4)  &3.54(3)  &2.15(6)  &6.5(1)   &  0.4(1)  & 100(10) &0.87(57) &2.57/0.95 \\
20083-01-03-020 &50721.19 &12580 &999  &0.99(7) &2.96(4)  &0.4(1)   &3.51(4)  &2.17(3)  &6.6(1)   &  0.28(9) & 100(10) &0.96(57) &2.56/0.95 \\
20083-01-04-02  &50722.25 &3360 &936  &0.99(6) &2.95(4)  &0.8(2)   &3.1(1)   &2.4(3)   &6.7(1)   &  0.3(1)  & 90(10) &0.73(57) &2.39/0.88 \\
20083-01-04-020 &50722.26 &16540 &1511 &0.99(6) &3.08(5)  &0.3(1)   &3.4(2)   &2.1(1)   &6.6(1)   &  0.3(1)  & 100(10) &0.98(57) &2.45/0.92 \\
30042-03-01-00  &51086.30 &8432 &1795 &0.99(2) &10.4(3)  &0.71(3)  &3.92(5)  &0.99(8)  &6.54(3)  &  0.61(1) & 50(20) &1.21(57) &2.62/3.67 \\ 
30042-03-02-00  &51093.21 &9712 &2421 &0.97(5) &3.16(4)  &0.41(5)  &5.35(8)  &3.34(4)  &6.53(2)  &  0.55(2) & 140(8) &1.29(57) &3.90/1.60 \\ 
30042-03-01-01  &51109.94 &1920 &1528 &0.94(3) &11.1(6)  &1.0(2)   &3.19(2)  &1.05(9)  &6.58(8)  &  0.53(7) & 30(20) &1.1(57) &2.19/3.44 \\
30042-03-02-01  &51110.01 &2352 &1525 &0.85(9) &9.6(2)   &0.52(4)  &3.52(3)  &0.93(2)  &6.53(4)  &  0.49(3) & 40(20) &0.96(57) &2.19/3.39 \\ 
30042-03-03-01  &51110.08 &4624 &1584 &0.89(9) &9.8(2)   &0.60(3)  &3.64(4)  &0.85(2)  &6.55(3)  &  0.49(2) & 40(20) &1.06(57) &2.28/3.48 \\ 
30042-03-04-00  &51113.01 &6608 &1583 &0.85(8) &9.6(1)   &0.50(2)  &3.85(2)  &1.12(3)  &6.55(2)  &  0.49(2) & 40(20) &1.07(57) &2.27/3.54 \\ 
30042-03-01-03  &51112.94 &2048 &1589 &0.86(9) &9.7(2)   &0.56(4)  &3.79(5)  &0.89(2)  &6.54(2)  &  0.52(3) & 60(20) &1.22(57) &2.29/3.56 \\ 
30042-03-01-04  &51115.07 &3312 &1599 &0.85(9) &9.9(2)   &0.53(3)  &3.68(2)  &0.94(3)  &6.53(2)  &  0.51(2) & 50(20) &1.07(57) &2.29/3.64 \\
30042-03-05-00  &51115.14 &6672 &1595 &0.88(7) &10.2(1)  &0.69(3)  &3.52(4)  &0.92(1)  &6.57(3)  &  0.54(2) & 70(20) &0.93(57) &2.28/3.60 \\ 
30042-03-07-01  &51119.94 &3376 &1719 &0.89(9) &9.6(2)   &0.53(2)  &4.03(3)  &0.9(2)   &6.50(1)  &  0.47(2) & 70(10) &0.92(57) &2.48/3.73 \\ 
30042-03-07-00  &51120.00 &6752 &1670 &0.88(6) &9.6(1)   &0.59(2)  &3.92(2)  &0.89(4)  &6.53(4)  &  0.51(2) & 70(20) &1.15(57) &2.40/3.66 \\ 
30042-03-08-00  &51122.17 &5296 &1637 &0.89(7) &10.10(2) &0.67(3)  &3.67(3)  &0.92(3)  &6.57(1)  &  0.55(2) & 50(30) &1.09(57) &2.40/3.66 \\ 
30042-03-10-00  &51127.77 &4432 &1787 &0.93(3) &8.2(2)   &0.46(2)  &4.35(3)  &1.19(3)  &6.51(2)  &  0.52(2) & 60(20) &1.3(57) &2.62/3.34 \\ 
30042-03-10-01  &51127.88 &2656 &1730 &0.92(2) &7.9(2)   &0.51(6)  &3.96(8)  &1.13(4)  &6.53(4)  &  0.69(9) & 50(30) &0.88(57) &2.54/3.17 \\ 
30042-03-11-00  &51127.93 &1013 &1752 &0.96(3) &8.4(3)   &0.55(3)  &4.07(3)  &1.15(2)  &6.46(2)  &  0.75(5) & 70(30) &0.95(57) &2.58/3.23 \\ 
30042-03-12-00  &51128.60 &16640 &1776 &0.98(2) &7.95(9)  &0.52(2)  &4.14(4)  &1.21(2)  &6.49(4)  &  0.69(4) & 60(30) &0.97(57) &2.63/3.08 \\ 
30042-03-13-00  &51128.93 &9824 &1765 &0.99(1) &8.13(9)  &0.53(4)  &4.09(2)  &1.23(2)  &6.47(3)  &  0.72(5) & 80(30) &0.76(57) &2.61/3.08 \\ 
30042-03-14-02  &51133.27 &880  &2059 &0.9(1)  &4.4(1)   &0.3(3)   &4.73(3)  &2.71(2)  &6.49(5)  &  0.57(7) & 160(20) &0.93(57) &3.23/1.88 \\ %#30042  FOREST PDS
30042-03-14-01  &51133.34 &2432 &2033 &1.00(9) &4.46(7)  &0.3(3)   &4.61(5)  &2.72(9)  &6.62(3)  &  0.54(4) & 150(20) &1.12(57) &3.18/1.04 \\ %#30042   FOREST PDS
30042-03-14-00  &51133.41 &2160 &2194 &0.99(6) &4.09(2)  &0.3(2)   &5.12(4)  &2.97(8)  &6.60(3)  &  0.62(4) & 160(20) &1.18(57) &3.46/1.90 \\ %#30042 alya liniin v PDS
30042-03-15-00  &51133.55 &13700 &1989 &1.00(3) &4.79(3)  &0.3(1)   &4.5(2)   &2.71(4)  &6.55(6)  &  0.55(2) & 120(20) &1.3(57)  &3.10/1.90 \\ 
30042-03-16-00  &51134.00 &14940 &1787 &1.00(5) &5.48(3)  &0.35(5)  &4.28(7)  &2.32(4)  &6.54(1)  &  0.55(2) & 120(20) &0.96(57) &3.75/2.10 \\ 
30042-03-17-00  &51134.54 &3200 &1686 &1.00(3) &6.9(2)   &0.40(8)  &3.91(6)  &1.79(7)  &6.54(4)  &  0.55(4) & 120(20) &1.2(57)  &3.75/2.10 \\
30042-03-18-00  &51193.32 &9488 &1232 &0.997(9)&14.4(3)  &1.07(1)  &2.58(3)  &0.97(6)  &6.49(2)  &  0.60(3) & 30(20) &1.12(57)  &1.78/2.79 \\ 
30042-03-19-01  &51195.20 &2432 &1253 &1.000(8)&14.3(5)  &1.01(1)  &2.47(4)  &0.95(9)  &6.53(2)  &  0.54(2) & 40(20) &1.15(57) &1.82/2.83 \\ 
30042-03-19-00  &51195.26 &9568 &1286 &0.99(3) &14.1(3)  &0.96(5)  &2.58(3)  &0.9(1)   &6.54(1)  &  0.55(1) & 50(20) &0.94(57) &1.86/2.90 \\ 
30042-03-20-00  &51196.94 &4608 &1365 &0.99(4) &12.06(4) &0.83(6)  &3.13(2)  &0.90(6)  &6.51(2)  &  0.56(2) & 50(10) &1.21(57)  &1.98/2.95 \\
40033-06-01-00  &51198.12 &10030 &1124 &0.99(7) &13.3(3)  &1.00$^{\dagger\dagger}$  &2.87(3)  &0.86(2)  &6.67(1)  &  0.30(4)&70(8) &1.29(58) &1.99/3.20 \\ 
40033-06-02-00  &51200.19 &9792 &1219 &0.99(5) &9.8(2)   &0.61(3)  &3.35(4)  &0.75(4)  &6.63(1)  &  0.43(4) & 60(10) &1.16(57) &2.20/2.94 \\ 
40033-06-02-01  &51201.91 &16420 &1293 &0.99(6) &7.93(8)  &0.44(2)  &3.89(2)  &1.04(3)  &6.64(2)  &  0.40(4) & 60(8) &1.04(57) &2.37/2.68 \\ 
40033-06-02-03  &51203.98 &13230 &1084 &1.00(3) &5.44(3)  &0.35(6)  &4.33(8)  &1.98(4)  &6.66(3)  &  0.36(1) & 40(10) &1.07(57) &2.64/2.08 \\
40033-06-02-04  &51205.92 &6576 &1685 &0.99(4) &5.43(5)  &0.33(7)  &4.3(1)   &1.96(5)  &6.65(2)  &  0.38(2) & 40(10) &0.96(57) &2.63/2.06 \\ 
40033-06-02-07  &51206.06 &2784 &1634 &0.99(2) &5.81(9)  &0.32(9)  &4.2(1)   &1.88(3)  &6.61(3)  &  0.35(3) & 40(10) &0.98(57) &2.54/2.07 \\
40033-06-02-05  &51206.12 &2704 &1716 &1.00(3) &5.37(6)  &0.4(1)   &4.3(2)   &1.73(8)  &6.64(3)  &  0.44(3) & 50(10) &0.87(57) &2.68/2.06 \\ 
40033-06-02-06  &51206.19 &3120 &996.9 &1.00(4) &5.55(8)  &0.4(1)   &3.8(1)   &1.89(3)  &6.69(5)  &  0.49(3) & 50(10) &1.19(57) &2.45/1.92 \\ 
40033-06-03-00  &51207.18 &3264 &1491 &1.00(2) &6.6(2)   &0.35(7)  &3.69(6)  &1.48(6)  &6.61(3)  &  0.39(3) & 30(10) &1.19(57) &2.30/2.12 \\ 
40033-06-03-06  &51207.27 &864  &911.7 &1.00(3) &6.8(3)   &0.8(1)   &3.5(1)   &1.33(9)  &6.58(6)  &  0.39(6) & 40(10) &1.2(57) &2.22/2.10 \\
40033-06-03-01  &51208.92 &6608 &1207 &0.99(2) &5.91(7)  &0.31(6)  &3.68(8)  &1.74(6)  &6.69(2)  &  0.33(2) & 30(10) &0.97(57) &2.28/1.88 \\ 
40033-06-03-02  &51210.12 &1856 &1635 &0.99(3) &4.27(3)  &0.24(6)  &3.81(5)  &2.16(6)  &6.68(3)  &  0.41(4) & 40(10) &1.16(57) &2.62/1.49 \\ 
40033-06-03-07  &51211.78 &2368 &1379 &1.00(2) &5.46(9)  &0.34(9)  &3.4(1)   &1.98(4)  &6.74(3)  &  0.35(3) & 30(10) &0.93(57) &2.15/1.64 \\ 
40033-06-03-03  &51211.91 &12350 &1112 &1.00(1) &5.71(5)  &0.30(6)  &3.37(7)  &1.54(3)  &6.68(1)  &  0.36(1) & 30(10) &1.37(57) &2.10/1.64 \\ 
40033-06-03-05  &51213.91 &5152 &1601 &0.99(3) &3.98(2)  &0.25(6)  &3.74(5)  &2.07(4)  &6.68(2)  &  0.39(2) & 30(10) &1.32(57) &2.58/1.59 \\
40027-06-01-01  &51237.04 &4528 &1023 &1.06(2) &7.8(3)   &0.59(2)  &2.38(4)  &0.78(2)  &6.56(4)  &  0.54(4) & 70(10) &1.17(57)  &1.58/1.57 \\
40027-06-01-02  &51237.18 &3360 &1087 &0.99(3) &7.3(2)   &0.38(8)  &2.69(6)  &0.89(4)  &6.57(3)  &  0.38(3) & 40(10) &0.74(57) &1.67/1.66 \\ 
40027-06-01-04  &51238.36 &3392 &851.8 &0.99(1) &7.9(3)   &0.39(7)  &2.42(4)  &0.84(2)  &6.51(3)  &  0.39(3) & 40(10) &1.09(57) &1.68/1.65 \\ 
40027-06-01-05  &51238.49 &2720 &1044 &0.91(9) &8.2(4)   &0.36(4)  &2.46(5)  &0.86(2)  &6.58(3)  &  0.41(3) & 50(10) &1.72(57)  &1.57/1.85 \\
40027-06-01-06  &51238.56 &2720 &1110 &0.97(4) &8.3(4)   &0.37(6)  &2.63(4)  &0.85(4)  &6.60(3)  &  0.32(3) & 40(10) &0.91(57) &1.68/1.88 \\ 
40027-06-01-03  &51238.76 &9616 &1054 &0.999(9)&9.6(2)   &0.43(3)  &2.45(3)  &0.73(3)  &6.61(1)  &  0.31(1) & 40(10) &0.95(57) &1.60/1.93 \\ 
40027-06-01-07  &51239.90 &3200 &657.9 &0.99(2) &8.9(3)   &0.41(7)  &2.41(3)  &0.85(6)  &6.62(3)  &  0.30(2) & 40(10) &1.13(57) &1.58/1.82 \\ 
40027-06-01-08  &51240.04 &1136 &1270 &1.00(4) &7.1(3)   &0.36(9)  &3.1(1)   &0.94(3)  &6.72(3)  &  0.21(3) & 40(10) &1.09(57) &1.95/1.99 \\ 
40027-08-01-02  &51359.15 &1056 &934.7 &1.00(4) &7.2(3)   &0.2(1)   &2.89(4)  &1.47(4)  &6.58(5)  &  0.29(4) & 40(10) &0.93(57) &1.81/1.69 \\ 
40019-03-02-02  &51409.17 &1984 &2078 &1.00(3) &3.02(3)  &0.52(8)  &5.7(2)   &3.16(6)  &6.64(3)  &  0.64(3) & 100(20) &1.37(57) &4.43/1.71 \\ 
40019-03-02-01  &51409.23 &2400 &2102 &0.94(6) &3.04(6)  &0.32(7)  &6.7(1)   &3.5(3)   &6.67(3)  &  0.39(4) & 150(10) &1.09(57) &4.47/1.71 \\
40019-03-02-00  &51409.55 &14370 &1268 &0.94(7) &2.94(2)  &0.43(5)  &7.54(9)  &3.1(2)   &6.64(2)  &  0.59(2) & 170(8) &1.12(57) &5.09/1.95 \\
40019-03-03-01  &51410.16 &2048 &1911 &0.96(8) &2.87(5)  &0.8(1)   &7.36(7)  &3.2(1)   &6.62(4)  &  0.44(2) & 160(10) &1.07(57) &5.20/2.18 \\
40019-03-03-00  &51410.23 &11760 &2321 &0.99(3) &3.02(2)  &0.40(4)  &7.21(6)  &3.16(4)  &6.62(1)  &  0.53(2) & 170(6) &1.05(57) &4.95/1.87 \\
40019-03-01-03  &51442.39 &8800 &438.6 &1.00(6) &5.7(3)   &0.3(1)   &2.6(1)   &2.7(6)   &6.64(6)  &  0.28(2) & 40(10) &0.83(57) &1.72/1.35 \\
40019-03-01-07  &51442.49 &1200 &457.6 &0.99(5) &5.6(2)   &0.3(1)   &2.6(1)   &2.6(3)   &6.75(6)  &  0.22(2) & 40(10) &1.09(57) &1.74/1.36 \\
40019-03-01-04  &51442.60 &1440 &692.9 &0.99(4) &5.3(2)   &0.3(1)   &2.7(1)   &2.1(4)   &6.73(5)  &  0.28(4) & 40(10) &1.53(57) &1.79/1.33 \\
40019-03-01-01  &51442.67 &1152 &705.3 &1.01(6) &5.2(1)   &0.4(1)   &3.0(1)   &2.7(6)   &6.69(5)  &  0.26(4) & 40(5) &1.17(57) &1.83/1.33 \\
40019-03-01-02  &51442.74 &848  &708.7 &1.00(4) &5.4(2)   &0.4(2)   &3.0(1)   &5.5(2)   &6.70(7)  &  0.21(4) & 40(8) &1.34(57) &1.84/1.36 \\
40019-03-01-00  &51442.80 &12610 &758.2 &0.99(3) &4.57(8)  &0.26(9)  &3.0(1)   &2.9(2)   &6.73(1)  &  0.26(4) & 40(10) &1.28(57) &1.99/1.25 \\
40019-03-01-06  &51443.19 &6288 &759.2 &0.99(2) &4.80(5)  &0.24(9)  &3.1(1)   &3.4(4)   &6.72(3)  &  0.23(2) & 40(10) &0.91(57) &2.06/1.33 \\
50023-01-01-00  &51610.70 &2656 &551.7 &0.95(3) &10.07(6) &0.63(6)  &4.04(5)  &0.93(5)  &6.56(3)  &  0.46(3) & 50(8) &1.03(57) &1.99/2.83 \\ 
50023-01-02-00  &51613.69 &2544 &801.5 &0.96(4) &10.4(2)  &1.00(1)  &2.93(4)  &0.85(4)  &6.62(3)  &  0.45(3) & 50(10) &1.52(57) &1.98/2.74 \\
50023-01-03-00  &51615.89 &3376 &788.8 &0.89(6) &8.5(2)   &0.40(2)  &3.21(3)  &0.9(3)   &6.56(3)  &  0.36(3) & 40(8) &1.75(57) &1.96/2.55 \\ 
50023-01-04-00  &51619.81 &3296 &520 &0.93(3) &7.8(3)   &0.42(4)  &2.96(4)  &1.07(5)  &6.58(3)  &  0.35(3) & 40(10) &1.07(57) &1.91/2.21 \\ 
50023-01-05-00  &51622.81 &3328 &773.5 &0.94(6) &6.7(3)   &0.22(2)  &3.49(3)  &3.69(4)  &5.99(7)  &  0.07(1) & 30(10) &0.76(57) &1.98/1.89 \\ 
50023-01-06-00  &51625.73 &2992 &536.6 &0.98(7) &5.5(2)   &0.32(5)  &3.11(5)  &1.5(2)   &6.58(4)  &  0.31(3) & 40(8) &1.04(57) &2.04/1.56 \\
50023-01-07-00  &51628.85 &2928 &1896 &0.97(9) &4.1(2)   &0.05(5)  &6.5(1)   &3.9(2)   &6.73(2)  &  0.72(3) & 150(6) &0.88(57) &4.01/2.02 \\
50029-23-01-00  &51652.4  &2256 &280.5 &1.00(3) &13.6(8)  &0.7(1)   &1.4(2)   &0.63(4)  &6.50(5)  &  0.29(3) & 40(10) &1.2(57) &1.03/1.44 \\ 
50029-23-01-01  &51652.46 &2224 &417.3 &0.89(6) &9.9(5)   &0.34(5)  &1.68(4)  &0.62(6)  &6.54(5)  &  0.19(3) & 30(10) &0.91(57) &1.04/1.42 \\ 
50029-23-02-00  &51657.13 &2464 &906.8 &0.89(3) &8.7(4)   &0.39(3)  &2.17(5)  &0.68(5)  &6.57(3)  &  0.27(3) & 40(10) &0.95(57) &1.38/1.77 \\  
50029-23-02-01  &51657.19 &2480 &1027 &0.97(6) &7.8(3)   &0.46(4)  &2.51(6)  &0.9(1)   &6.63(3)  &  0.26(3) & 40(8) &1.29(57) &1.57/1.80 \\  
50029-23-02-02  &51657.58 &5664 &773.8 &0.96(2) &7.9(2)   &0.38(2)  &2.49(3)  &1.2(2)   &6.62(2)  &  0.27(2) & 40(10) &0.94(57) &1.52/1.72 \\  
50029-23-02-03  &51657.73 &7408 &384.3 &0.99(8) &8.6(2)   &0.40(4)  &2.28(4)  &1.3(3)   &6.58(2)  &  0.26(2) & 40(10) &0.81(57) &1.43/1.64 \\  
50023-01-12-00  &51663.96 &1792 &1195 &0.99(3) &5.5(1)   &0.3(1)   &2.9(1)   &2.6(4)   &6.75(4)  &  0.28(3) & 40(10) &1.2(57) &1.90/1.43 \\  
50023-01-13-00  &51667.55 &2592 &1715 &1.02(4) &2.95(4)  &0.48(9)  &3.8(1)   &2.3(3)   &6.64(3)  &  0.36(3) & 50(8) &1.1(57) &2.88/1.08 \\ 
50023-01-14-00  &51669.55 &2528 &961.9 &1.0(2)  &4.07(9)  &0.2(4)   &3.07(6)  &2.95(9)  &6.72(5)  &  0.27(3) & 40(6) &1.24(57) &2.01/1.06 \\ 
50023-01-15-00  &51673.61 &2048 &313.9 &1.00(4) &10.1(6)  &0.47(9)  &1.73(3)  &0.74(4)  &6.51(4)  &  0.25(4) & 40(10) &1.09(57) &1.16/1.42 \\ 
50023-01-17-00  &51679.72 &2688 &368.7 &1.01(6) &9.4(4)   &0.6(1)   &2.15(4)  &0.40(5)  &6.55(4)  &  0.32(3) & 50(10) &0.78(57) &1.41/1.74 \\ 
50023-01-18-00  &51682.51 &2656 &372.2 &0.89(6) &7.9(3)   &0.32(3)  &2.61(3)  &1.5(4)   &6.63(5)  &  0.23(4) & 40(10) &0.92(57) &1.55/1.86 \\ 
50023-01-19-00  &51685.50 &2720 &632.4 &0.97(5) &7.3(2)   &0.35(3)  &2.8(3)   &1.9(2)   &6.70(4)  &  0.26(3) & 40(6) &0.81(57) &1.72/1.76 \\ 
50023-01-20-00  &51688.76 &2608 &613.7 &0.94(4) &8.1(3)   &0.38(3)  &2.67(5)  &1.0(3)   &6.57(4)  &  0.28(3) & 50(8) &0.92(57) &1.63/1.92 \\ 
50023-01-21-00  &51691.71 &3056 &654.4 &1.01(9) &9.4(4)   &0.62(7)  &2.80(3)  &0.58(7)  &6.56(3)  &  0.32(3) & 60(10) &1.19(57) &1.73/2.19 \\ 
50023-01-22-00  &51695.33 &2720 &476.1 &1.01(9) &8.6(2)   &0.56(6)  &3.02(4)  &0.81(3)  &6.59(4)  &  0.31(4) & 60(10) &1.1(57) &1.82/2.22 \\ 
50023-01-23-00  &51697.45 &2112 &248 &0.99(2) &7.8(3)   &0.52(8)  &2.89(6)  &0.92(4)  &6.68(6)  &  0.23(5) & 40(10) &0.89(57) &1.83/2.14 \\ 
50023-01-24-00  &51709.69 &2560 &188.5 &0.97(4) &9.5(6)   &0.7(2)   &1.96(3)  &0.5(2)   &6.64(5)  &  0.30(4) & 40(10) &1.29(57) &1.38/1.83 \\ 
50023-01-25-00  &51712.35 &2672 &803.3 &0.90(4) &7.8(2)   &0.30(1)  &2.63(4)  &1.8(5)   &6.72(3)  &  0.19(2) & 30(10) &1.09(57) &1.56/1.84 \\ 
50023-01-26-00  &51715.41 &2528 &576.4 &0.99(7) &13.8(7)  &0.51(4)  &2.36(3)  &0.8(3)   &6.59(4)  &  0.16(2) & 30(10) &1.19(57) &1.52/2.26 \\ 
50023-01-27-00  &51718.93 &3024 &655.6 &0.99(3) &9.9(5)   &0.7(1)   &2.58(5)  &0.61(6)  &6.67(3)  &  0.24(2) & 40(10) &0.95(57) &1.72/2.28 \\ 
50023-01-28-00  &51721.86 &2880 &238.6 &1.00(2) &9.3(3)   &0.6(1)   &2.54(4)  &0.61(4)  &6.56(4)  &  0.35(2) & 50(6) &1.07(57) &1.74/2.23 \\ 
50023-01-29-00  &51724.57 &3344 &1063 &0.99(3) &5.92(9)  &0.27(9)  &3.4(1)   &1.7(2)   &6.57(3)  &  0.28(2) & 50(10) &1.1(57) &2.13/1.71 \\ 
50023-01-30-00  &51727.50 &3088 &950.2 &0.99(2) &6.4(2)   &0.24(7)  &3.1(1)   &2.1(3)   &6.71(3)  &  0.16(2) & 30(10) &1.18(57) &1.90/1.62 \\ 
50023-01-31-00  &51730.42 &2768 &226 &0.99(4) &6.7(3)   &0.3(1)   &2.62(8)  &0.9(1)   &6.58(5)  &  0.28(5) & 40(10) &1.02(57) &1.71/1.59 \\ 
50023-01-32-00  &51733.41 &2912 &817.7 &1.0(2)  &3.7(1)   &0.2(1)   &3.4(1)   &1.9(2)   &6.61(4)  &  0.35(4) & 50(8) &1.05(57) &1.36/1.11 \\ 
      \enddata%     \hline%      \end{tabular}
    \label{tab:fit_table}
%  \end{center}
$^\dagger$ The spectral model is  $wabs*(blackbody + COMPTB + Gaussian)$, where $N_H$ is fixed at a value 2.73$\times 10^{22}$ cm$^{-2}$ \citep{piraino00};
% (Piraino et al., 2000); 
color temperature $T_s$ and $T_{BB}$ are fixed at 1.3 and 0.7 keV, respectively (see comments in the text); 
$^{\dagger\dagger}$ when parameter $\log(A)\gg1$, it is fixed at 1.0 (see comments in the text), 
$^{\dagger\dagger\dagger}$ normalization parameters of $blackbody$ and $COMPTB$ components are in units of 
$L_{37}/d^2_{10}$ $erg/s/kpc^2$, where $L_{37}$ is the source luminosity in units of 10$^{37}$ erg/s, 
$d_{10}$ is the distance to the source in units of 10 kpc 
and $Gaussian$ component is in units of $10^{-2}\times total~~photons$ $cm^{-2}s^{-1}$ in line %({\bf Lenochka please specify units for these components, look at XSPEC manual})
,  
%$\sigma_{line2}$ of Gaussian2 component is fixed to a value 0.01 keV (see comments in the text), 
$^{\dagger\dagger\dagger\dagger}$spectral fluxes (F$_1$/F$_2$) in units of $\times 10^{-9}$ ergs/s/cm$^2$ for  (3 -- 10) and (10 -- 60) keV energy ranges respectively.  
%in units of $\times 10^{-9}$ ergs/s/cm$^2$.
%* this observations are  fitted with $bmc+Gaussian1+Gaussian2+bbody$ model, see values of the best-fit BB color temperature and EW in Table 2, 3 and 4.
\end{deluxetable}
~~~~

\newpage

\begin{figure}[ptbptbptb]
\includegraphics[scale=0.9,angle=0]{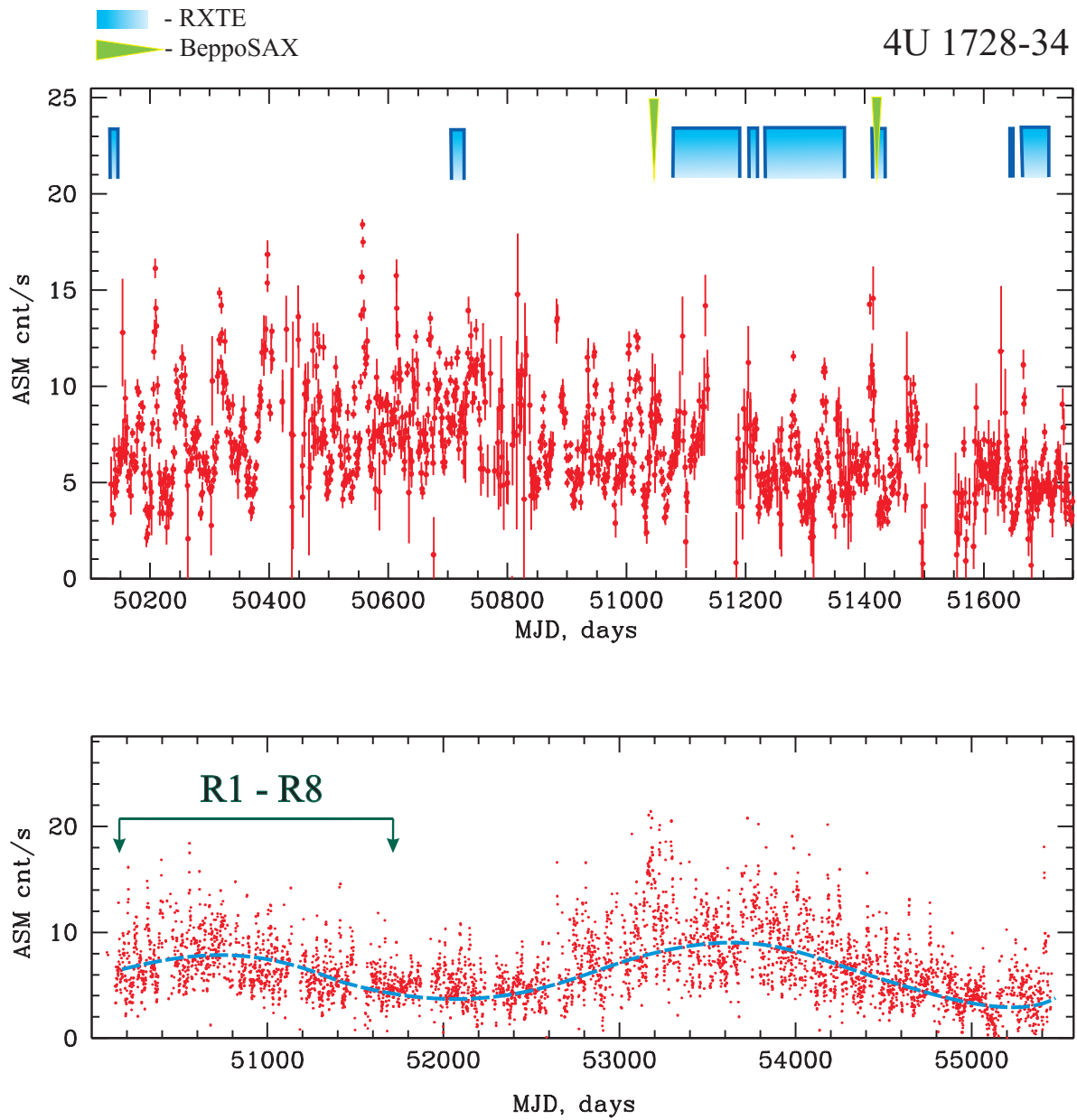}
\caption{
{\it Top:}  
Evolution of ASM/RXTE count rate %, flux density $S_{8.46 GHz}$ at 8.46 GHz (VLA), 
%BMC normalization and 
%photon index $\Gamma$ 
during  1996 -- 2006 observations of 4U~1728-34. Blue rectangles  indicate 
the RXTE/PCA \& HEXTE data of 
pointed observations, 
green triangles show $Beppo$SAX NFI data used for the analysis.
{\it Bottom:}  
ASM/RXTE 1-day type light curve of 4U~1728-34 during 1996 -- 2010. 
Blue dashed line shows  a mean count rate %of %of  1-day type ASM light curve 
and indicates the long-term quasi-periodic variability of mean soft flux 
during  $\sim$ six years cycle. Green double arrow points out the 1996 -- 2000 time interval 
of {\it RXTE} observations, used in our analysis (see R1 -- R8 intervals).
%Red triangles/black circles ({\it for two last panels}) correspond 
%to hard/soft components with $\Gamma_1$/$\Gamma_2$, respectively. 
%{\it Bottom:} Spectral index $\Gamma$ plotted versus BMC normalization ({\it left}) 
%and Comptonized fraction ({\it right}) for this rise transition. Here red 
%triangles/black circles correspond to hard/soft components with 
%$\Gamma_1$/$\Gamma_2$, correspondingly.
}
\label{outburst_05_rise}
\end{figure}

\newpage
\begin{figure}[ptbptbptb]
\includegraphics[scale=1.2, angle=0]{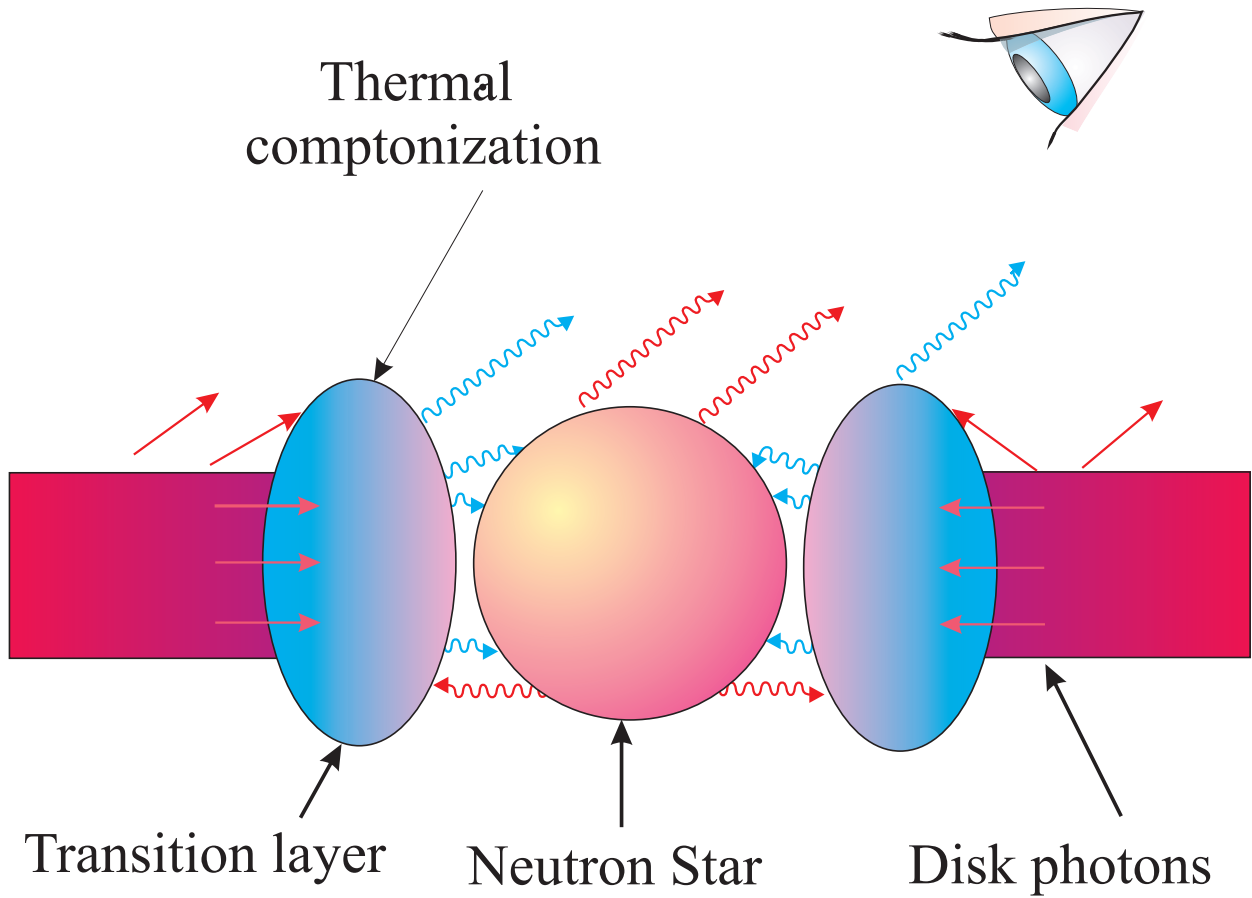}
\caption{A suggested  geometry of the system.   Disk and neutron star soft photons are 
up-scattered 
(Comptonized) in the relatively hot plasma of the transition layer 
({\it between the accretion disk and NS surface}).  But some fraction of these 
photons is seen directly by the Earth observer. Red and blue photon trajectories correspond to soft 
and hard (upscattered) photons respectively.
}
\label{geometry}
\end{figure}

\newpage 
\begin{figure}[ptbptbptb]
\includegraphics[scale=0.9,angle=0]{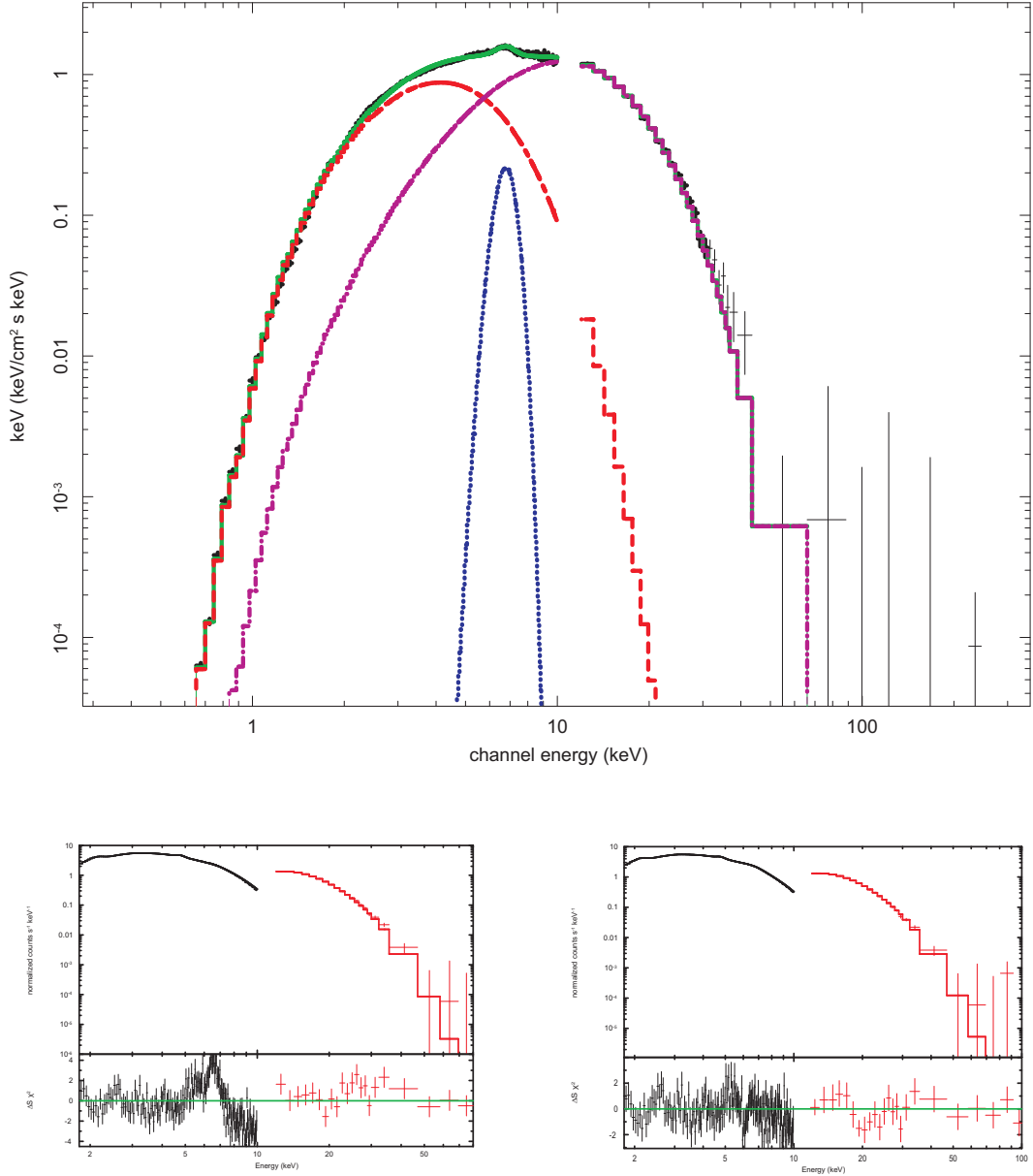}
\caption{$Top:$  the best-fit spectrum of 4U~1728-34 %during {\it Intermediate state} events 
 in $E*F(E)$ units using {\it Beppo}SAX observation  20889003 carried out on 19th of  April 1999.   The data are presented by crosses and the best-fit spectral  model   {\it wabs*(blackbody+COMPTB+Gaussian)} by green line. The model components  are shown by  red, crimson and  blue lines for {\it blackbdody}, {\it COMPTB}  and {\it Gaussian} components respectively. 
 $Bottom~panels$: the spectrum in units of counts along  with $\Delta \chi$.
Left {\it bottom panel} : the best-fit spectrum and 
 $\Delta \chi$ for the model fit without the line component (reduced $\chi^2$=2.15 for 445 d.o.f) and on 
{\it right bottom panel}:  same as that  on the {\it left} one   but with  an addition of  {\it Gaussian}  
(K$_{\alpha}$-line) component 
 (reduced $\chi^2$=1.16 for 445 d.o.f). The best-fit model parameters are 
$\Gamma$=2.07$\pm$0.04, $T_e$=3.29$\pm$0.04 keV, $E_{line}$=6.0$\pm$0.1 keV  and EW$_{line}=51\pm 11$ eV
(see more details in Table 3).
%On the top panel %are denoted by green points,
% spectral model {\it wabs*(blackbody+compTB+Gaussian)} 
%presented with components are shown by red, crimson and  blue lines for {\it Bbdody}, {\it CompTB} 
%and {\it Gaussian} components respectively.
}
\label{BeppoSAX_spectra}
\end{figure}

\newpage
\begin{figure}[ptbptbptb]
\includegraphics[scale=0.85,angle=0]{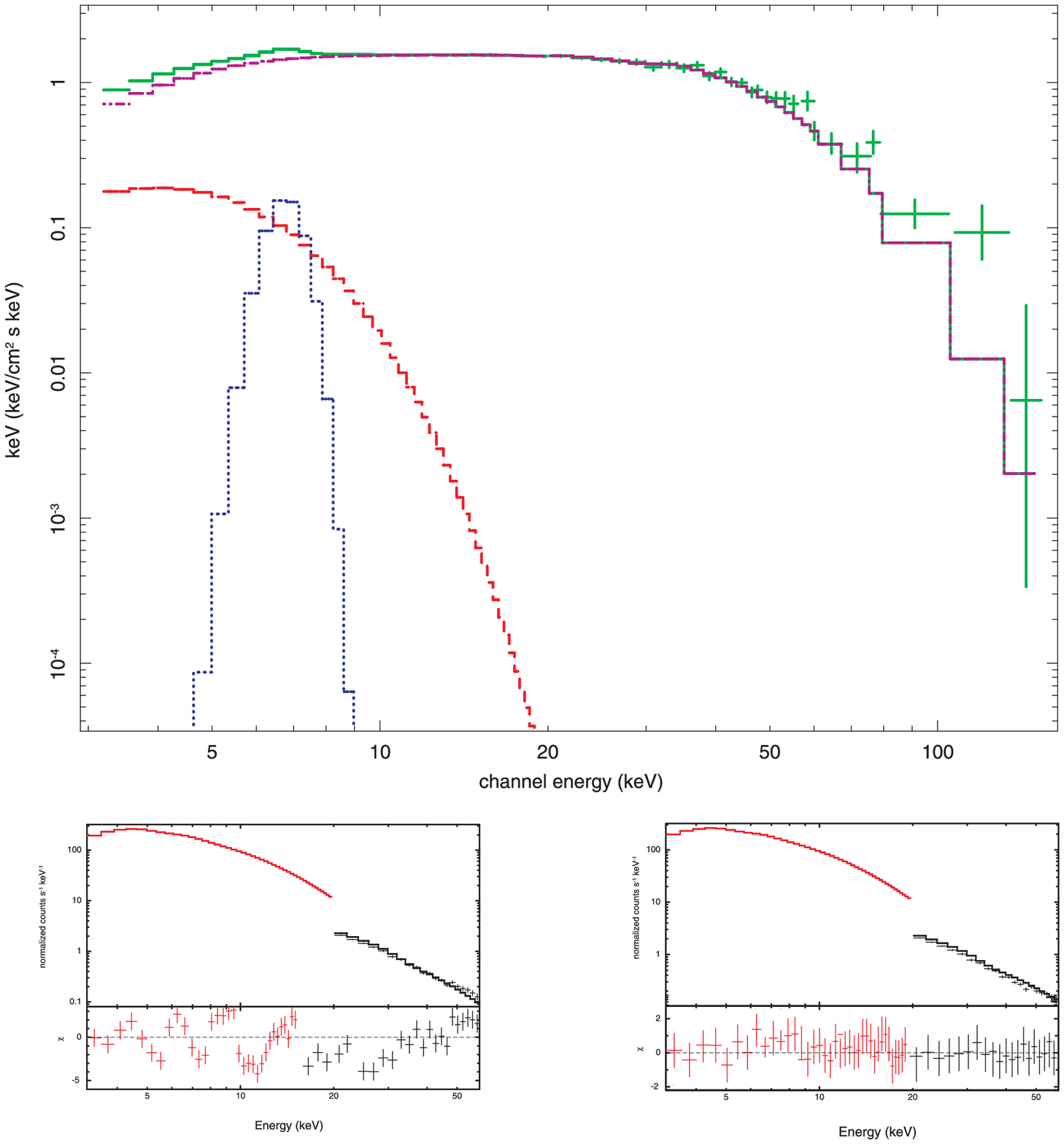}
\caption{The best-fit {\it RXTE} spectrum of 4U~1728-34 for the  low luminosity state in units 
$E*F(E)$ ($top$) and the spectra in
% normalised 
counts  units ({\it bottom panels}) with 
$\Delta\chi$ for the 30042-03-01-00 observation. {\it Left bottom panel}: 
a  fit  of the model $wabs*COMPTB$, 
%withoutt the line at $\sim$6.7~keV 
%and the $Blackbody$  component at low energies 
($\chi^2_{red}$=2.1 for 61 d.o.f.) and 
{\it right bottom panel}: the same as the latter one but with addition of an iron $Gaussian$ line and 
the $blackbody$ component,  namely using the model $wabs*(blackbody+COMPTB+Gaussian)$  ($\chi^2_{red}$=1.18 for 57 dof). 
The best-fit model parameters are 
$\Gamma$=1.99$\pm$0.02, $T_e$=10.4$\pm$0.3 keV and $E_{line}$=6.54$\pm$0.03 keV 
(see more details in Table 4). Red, violet and blue lines stand 
$blackbody$, $COMPTB$ and $Gaussian$ components respectively.
%A typical $E*F(E)$ spectral diagram  of 4U~1728-34 during hard (low) state event of 1996 
%for {\it RXTE}/PCA/HEXTE observation 30042-03-01-00.
}
\label{rxte_hard_state_spectrum}
\end{figure}

\newpage
\begin{figure}[ptbptbptb]
\includegraphics[scale=0.9,angle=0]{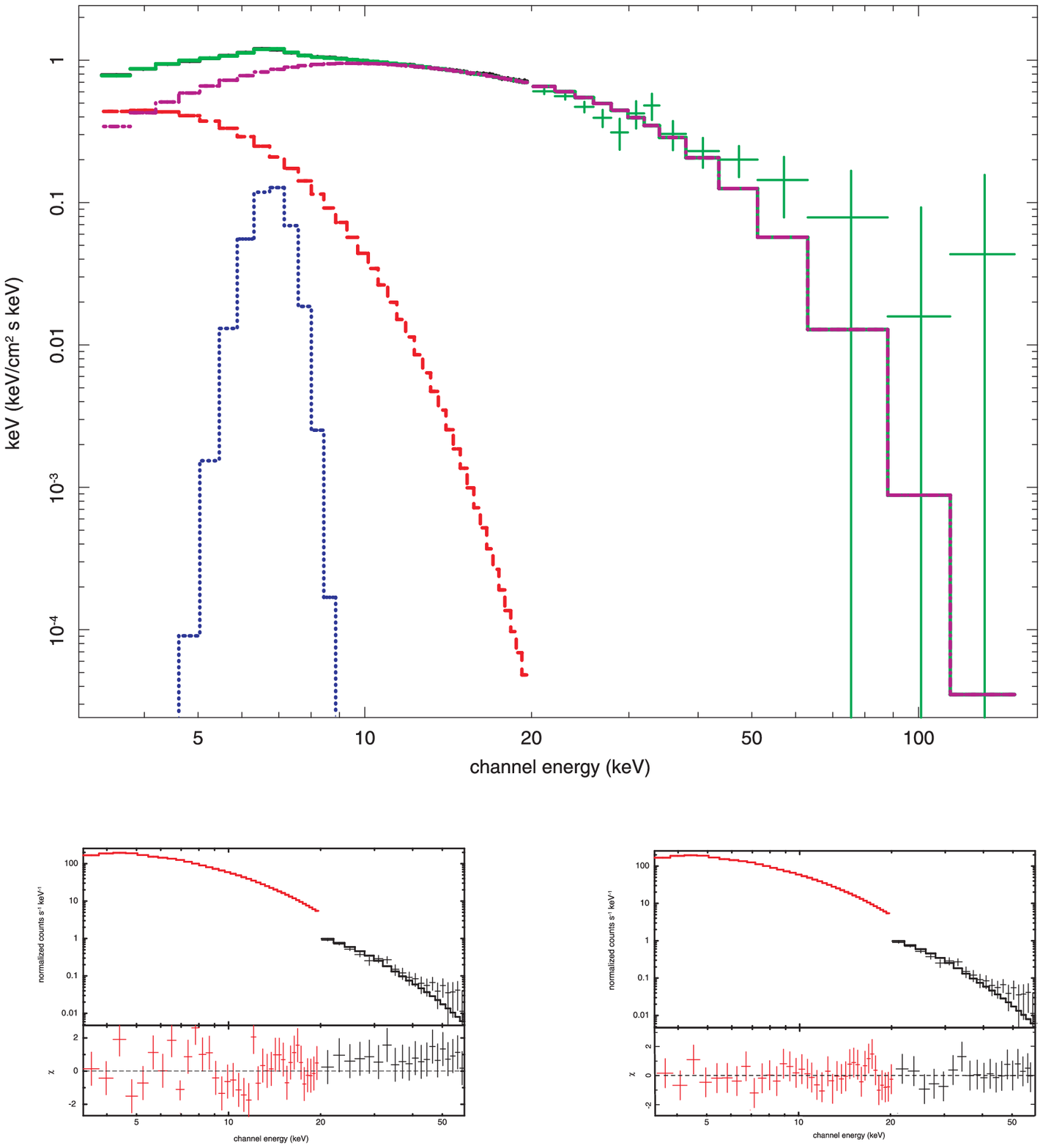}
\caption{The best-fit {\it RXTE} spectrum of 4U~1728-34 for  the high luminosity state in units
$E*F(E)$ ($top$) and the spectra in 
%normalized 
counts  units ({\it bottom panel}) with 
$\Delta\chi$ for the 50023-01-12-00 observation. {\it Left bottom panel}: 
a fit of the model $wabs*CompTB$
%, without modeling the line at $\sim$6.7 keV 
%and the $Blackbody$  component at low energies 
($\chi^2_{red}$=1.79 for 61 dof) and 
{\it right bottom panel}: the same as the latter one but with addition  of an iron $Gaussian$ line and 
the $Blackbody$ component, namely  using the model 
$wabs*(blackbody+COMPTB+Gaussian)$ 
($\chi^2_{red}$=1.2 for 57 dof). The best-fit model parameters are 
$\Gamma$=1.99$\pm$0.03, $T_e$=5.5$\pm$0.1 keV and $E_{line}$=6.75$\pm$0.04 keV 
(see more details  in Table 4). Red, violet and blue lines stand for  
$Blackbody$, $COMPTB$ and $Gaussian$ components, respectively.
%A typical $E*F(E)$ spectral diagram  of 4U~1728-34 during soft (high) state event of 2000 
%for {\it RXTE}/PCA/HEXTE observation 50023-01-12-00. 
}
\label{rxte_soft_state_spectrum}
\end{figure}

\newpage
\begin{figure}[ptbptbptb]
\includegraphics[scale=0.9,angle=0]{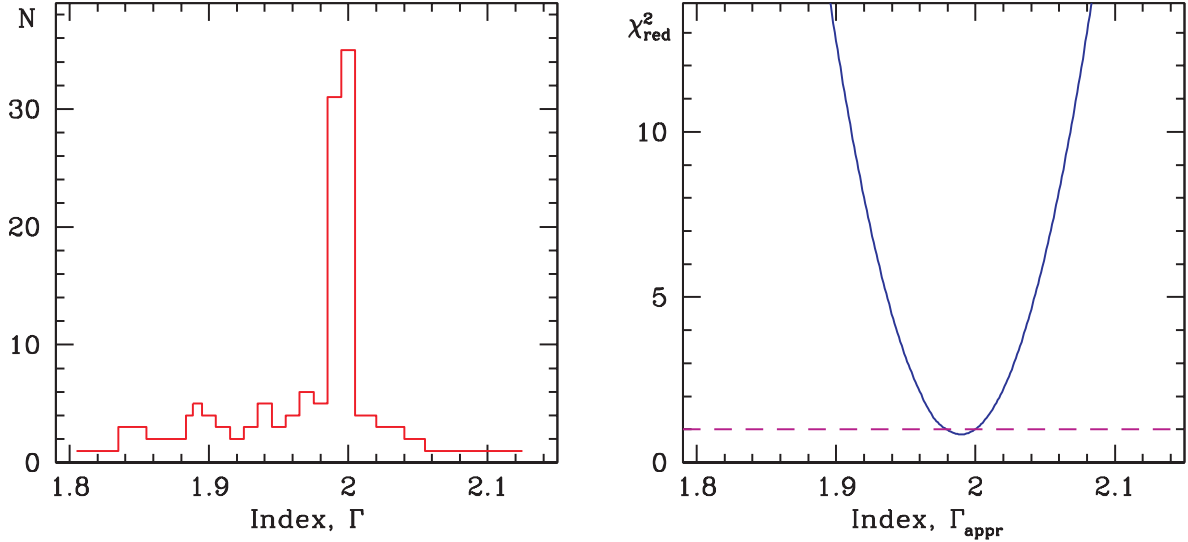}
\caption{{\it Left panel:} Histogram (frequency distribution) of the best-fit photon power-law 
index $\Gamma$  obtained using a model $wabs*(blackbody+COMPTB+Gaussian)$  for  {\it RXTE} data (1996 -- 2000). 
%of 4U~1728-34. % (see comments in the text). 
{\it Right panel:} Function  $\chi^2(\Gamma_{appr})=\frac{1}{N}\sum_{i=1}^N\left(
\frac{\Gamma_i-\Gamma_{appr}}{\Delta\Gamma_i}
\right)^2$ vs $\Gamma_{appr}$. A dashed horisontal line indicates the critical 
residual level $\chi^2_{red}=1$ (see comments in the text). 
}
\label{hist}
\end{figure}

\newpage
\begin{figure}[ptbptbptb]
\includegraphics[scale=0.9,angle=0]{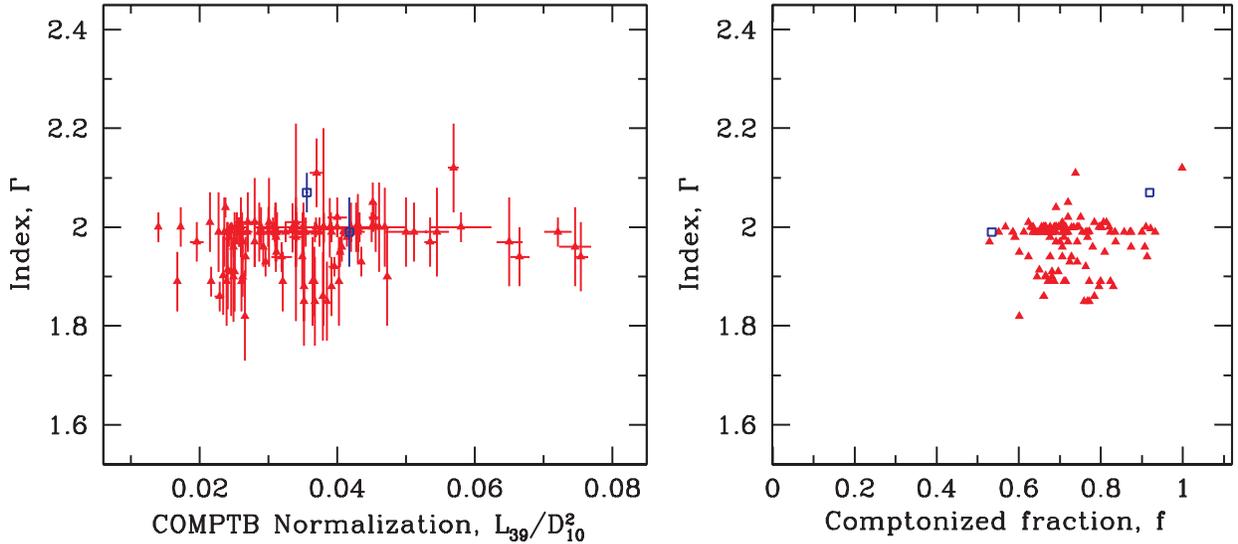}
\caption{
{\it On the left}  photon index $\Gamma$ is  plotted versus  COMPTB normalization 
$N_{COMPTB}=L_{39}/D^2_{10}$ and  {\it on the  right} that versus
Comptonized fraction $f=A/(1+A)$    using  our spectral model 
$wabs*(blackbody+COMPTB+Gaussian)$ 
%during flare spectral  transitions 
(see  details in Tables 3, 4). 
 Blue and   red points correspond to {\it Beppo}SAX  and {\it RXTE} observations of 4U~1728-34 
respectively. 
}
\label{outburst_index_norm}
\end{figure}

\newpage
\begin{figure}[ptbptbptb]
\includegraphics[scale=0.95,angle=0]{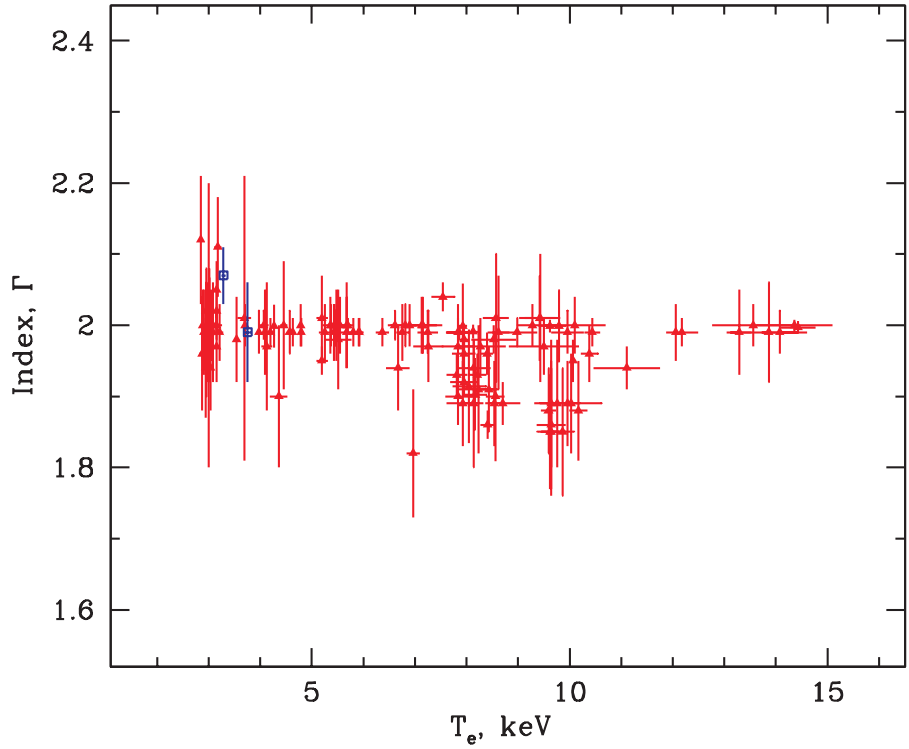}
\caption{
Photon index plotted versus electron temperature $T_e$ (in keV)  in the framework of  our spectral model $wabs*(blackbody+COMPTB+Gaussian)$ 
%during flare transitions 
(see details in Tables 3, 4). 
 Blue and   red points correspond to {\it Beppo}SAX  and {\it RXTE} observations of 4U~1728-34 
respectively. 
}
\label{outburst_index_temperature}
\end{figure}

\newpage
\begin{figure}[ptbptbptb]
\includegraphics[scale=0.8,angle=0]{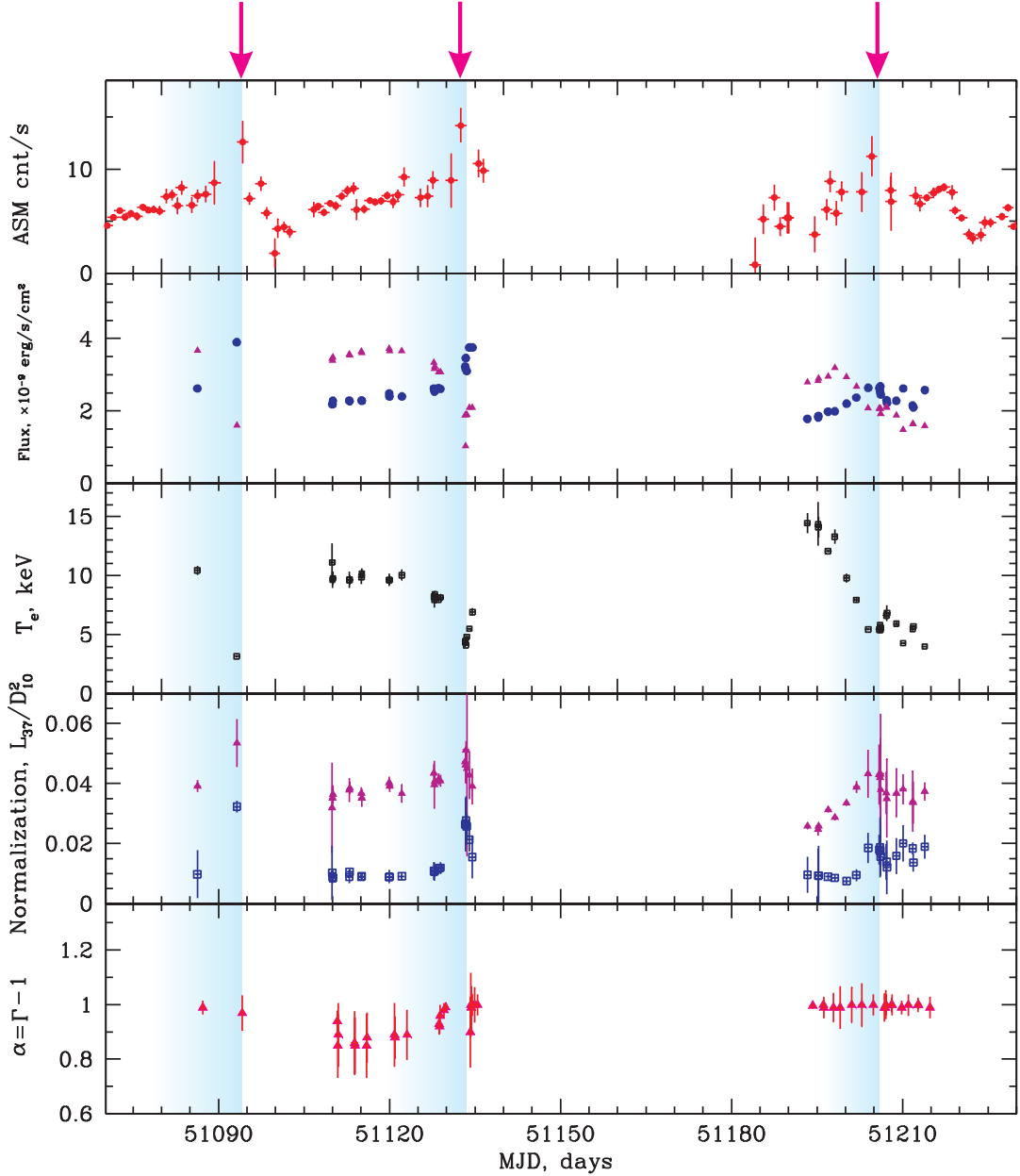}
\caption{
{\it From Top to Bottom:}
Evolution  of the {\it RXTE}/ASM count rate, model flux in 3-10 keV  and 10-60 keV energy ranges 
({\it blue and crimson} points respectively), electron temperature $kT_e$ in keV 
 and $COMPTB$ and $blackbody$ %and $Gaussian$
 normalizations 
({\it crimson and blue} %, green} 
respectively)  during 1998, 1999 flare transition set ({\it R3, R5}). 
In the last bottom panel we present   an evolution of the spectral index $\alpha=\Gamma-1$.
% as a function of time.
%Blue, orange and pink vertical  
%dashed lines indicate moments at MJD = 51128, 51133, and 51134 (before, during, and  after 
%X-ray flare, respectively), which used for spectral and temporal analysis as A, B and C points in Fig.~\ref{ev_PDS_SP}.  
The rising phases  of the burst are marked with blue vertical strips. 
The peak burst times are indicated by the arrows on the top of the plot.
% (see Fig.~\ref{ev_PDS_SP}). 
}
\label{lc_1998}
\end{figure}

\newpage
\begin{figure}[ptbptbptb]
\includegraphics[scale=0.8,angle=0]{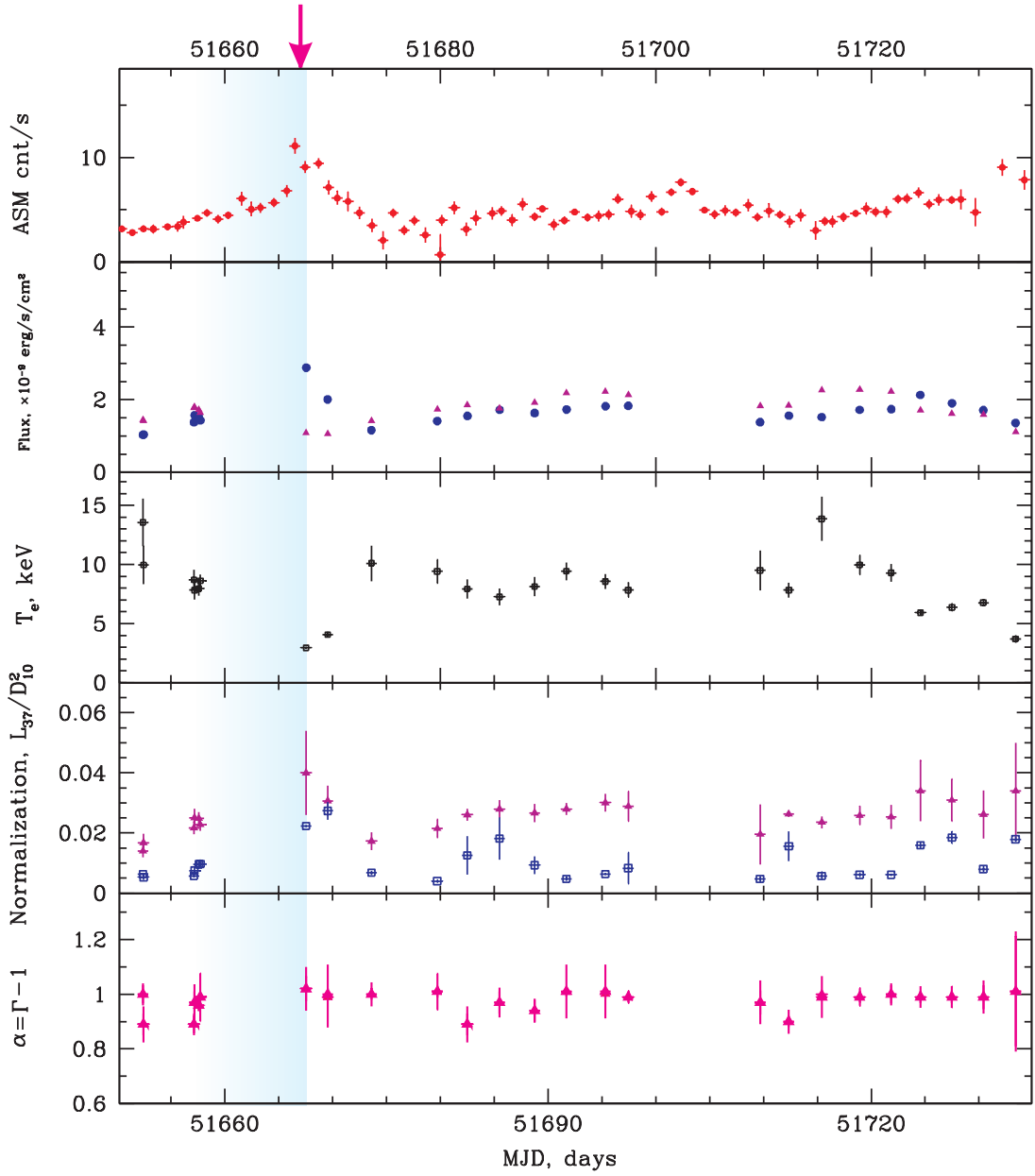}
\caption{ Similar to  that presented in Fig. \ref{lc_1998} but  for the {\it RXTE} 2000 flare transition set  ({\it R7, R8}).
%{\it From Top to Bottom:}
%Evolutions of the RXTE/ASM count rate, model flux at [3-10]/[10-60] keV 
%({\it blue/crimson}), electon temperature and $CompTB$/$Bbody$/$Gauss$ normalization 
%({\it crimson/blue/green}) 
%during 2000 outburst transition set 
}
\label{evolution_lc_3}
\end{figure}

\newpage
\begin{figure}[ptbptbptb]
\includegraphics[scale=1,angle=0]{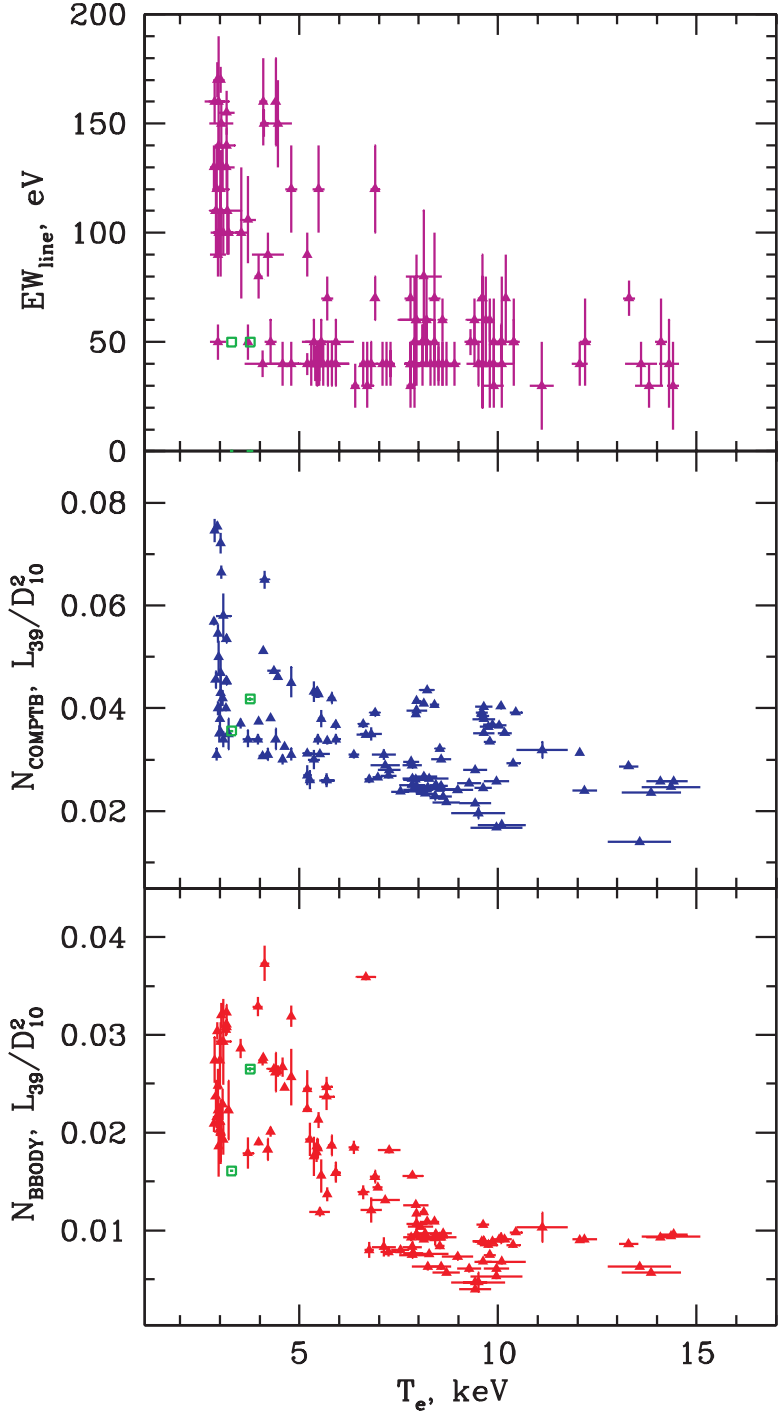}
\caption{Equivalent width of the iron line (in eV, $top$), 
normalizations of $COMPTB$ ($middle$) and $Blackbody$ ($bottom$) components 
plotted versus electron temperature $T_e$ (in keV)  
in the framework  of  our spectral model 
$wabs*(blackbody+COMPTB+Gaussian)$ during flare transitions (see also Tables 3, 4). 
Green points correspond to {\it Beppo}SAX observations of 4U~1728-34. 
% Blue/green and red/violet (top/bottom) points correspond to {\it Beppo}SAX  and {\it RXTE} observations of 4U~1728-34 
%respectively. 
}
\label{EW_norm_temperature}
\end{figure}

\newpage
\begin{figure}[ptbptbptb]
\includegraphics[scale=0.8,angle=0]{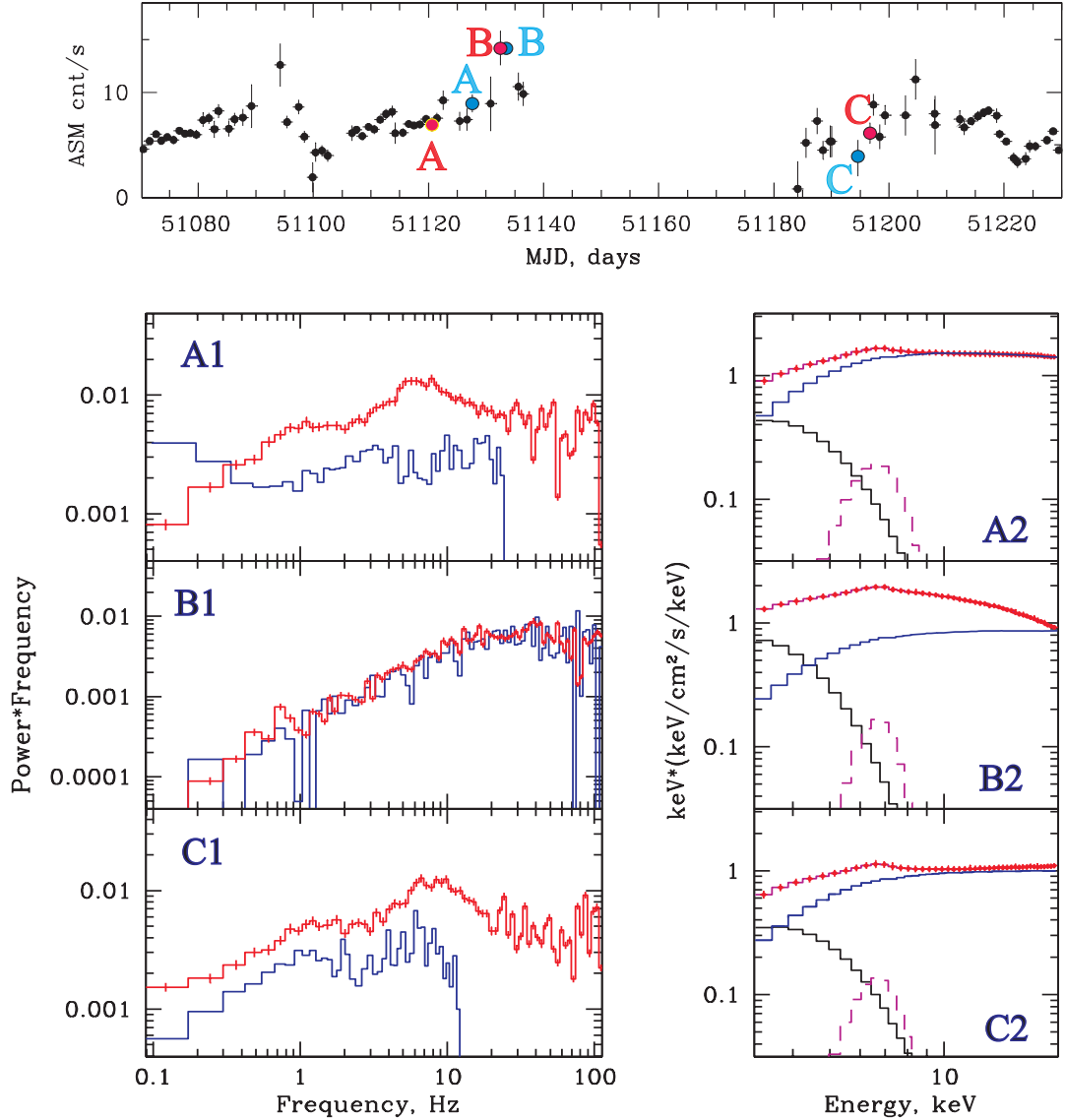}
\caption{
{\it Upper} panel:  RXTE/ASM count rate during the 1998 -- 1999 ($R3$, $R5$) outburst transition. Red/blue points  A, B, and C mark moments at MJD = 51122/51128, 51133.27/51133.34 and 51196/51193 (before, during, 
and after X-ray outburst, respectively).  
 {\it Lower} panels: PDSs for 13-30 keV energy band ($left$ column) are plotted along with energy spectral diagram $E*F(E)$ ($right$ column) related to A , B  and C points  of X-ray light curve
(upper panel).  
%$Red$ histograms corresponds to aforementioned MJD moments, whereas $blue$ histograms refer to adjacent observations  to illustrate continuous fast evolution of PDSs. 
The strong noise component related to break at 1 -- 3 Hz and broad QPOs centered at 7 -- 10 Hz are present  before and after burst (see panels A1, C1).   At the X-ray flare peak (see panel B1) one can see a white-red noise PDS with the break at  about 40 Hz.   
On the {\it right}  hand side panels we present the $E*F(E)$ spectral diagrams (panels A2, B2, C2) related to the corresponding power spectra 
(panels A1, B1, C1).  The data are shown by red points and  
 the spectral model components are displayed  by blue, black, and dashed purple lines for $COMPTB$, $blackbody$ and $Gaussian$ components respectively.
}
\label{ev_PDS_SP}
\end{figure}

\newpage
\begin{figure}[ptbptbptb]
\includegraphics[scale=0.9,angle=0]{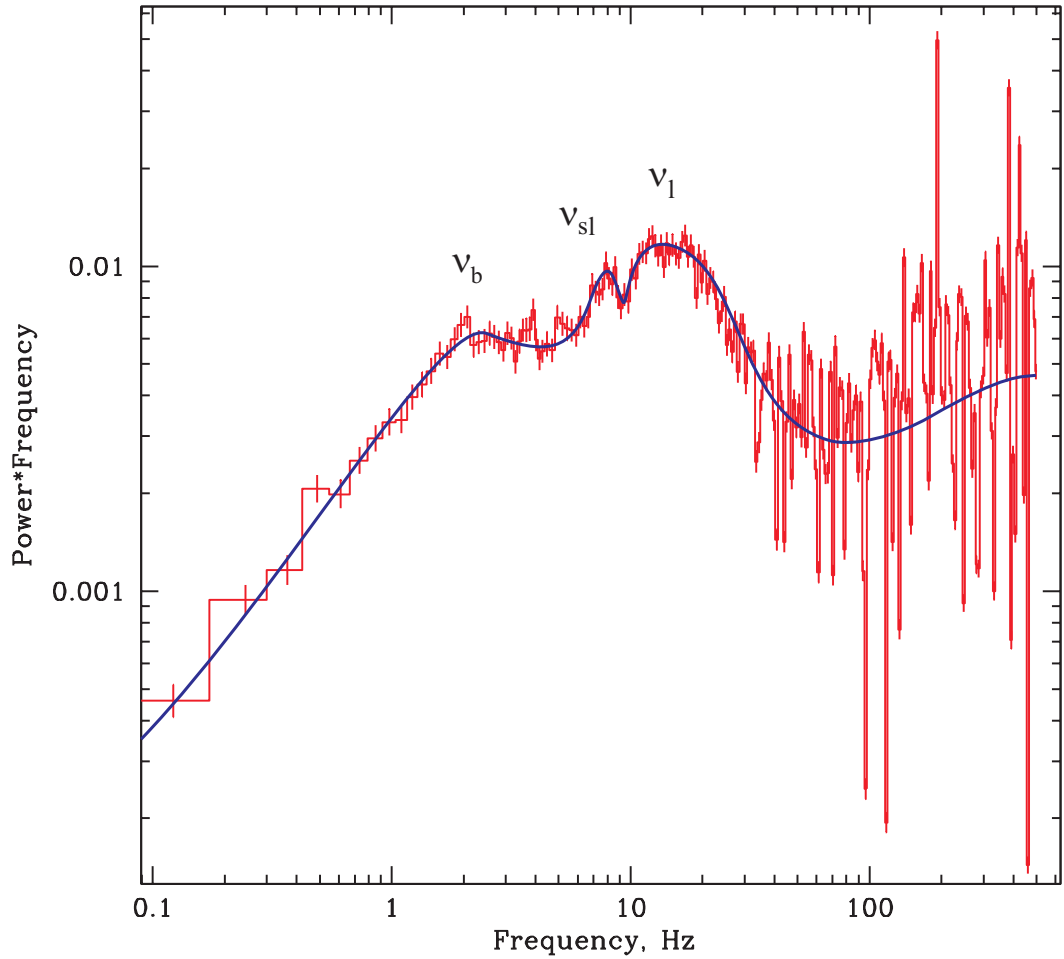}
\caption{The   $\nu\times power$ diagram of 4U~1728-34 in 0.1 -- 150 Hz range,  observed on March 7, 2000 
(50023-01-01-00, MJD=51610). Blue solid line shows the best fit of the  power spectrum which  
typically consists of three components: the broad-band noise with break $\nu_b$ (broken power law), low frequency QPOs fit by Lorentzians 
($\nu_{sl}$, $\nu_l$) and $\sim$100 Hz bump noise [see  \cite{disalvo2001}].
}
\label{PDS}
\end{figure}

\newpage
\begin{figure}[ptbptbptb]
\includegraphics[scale=0.90,angle=0]{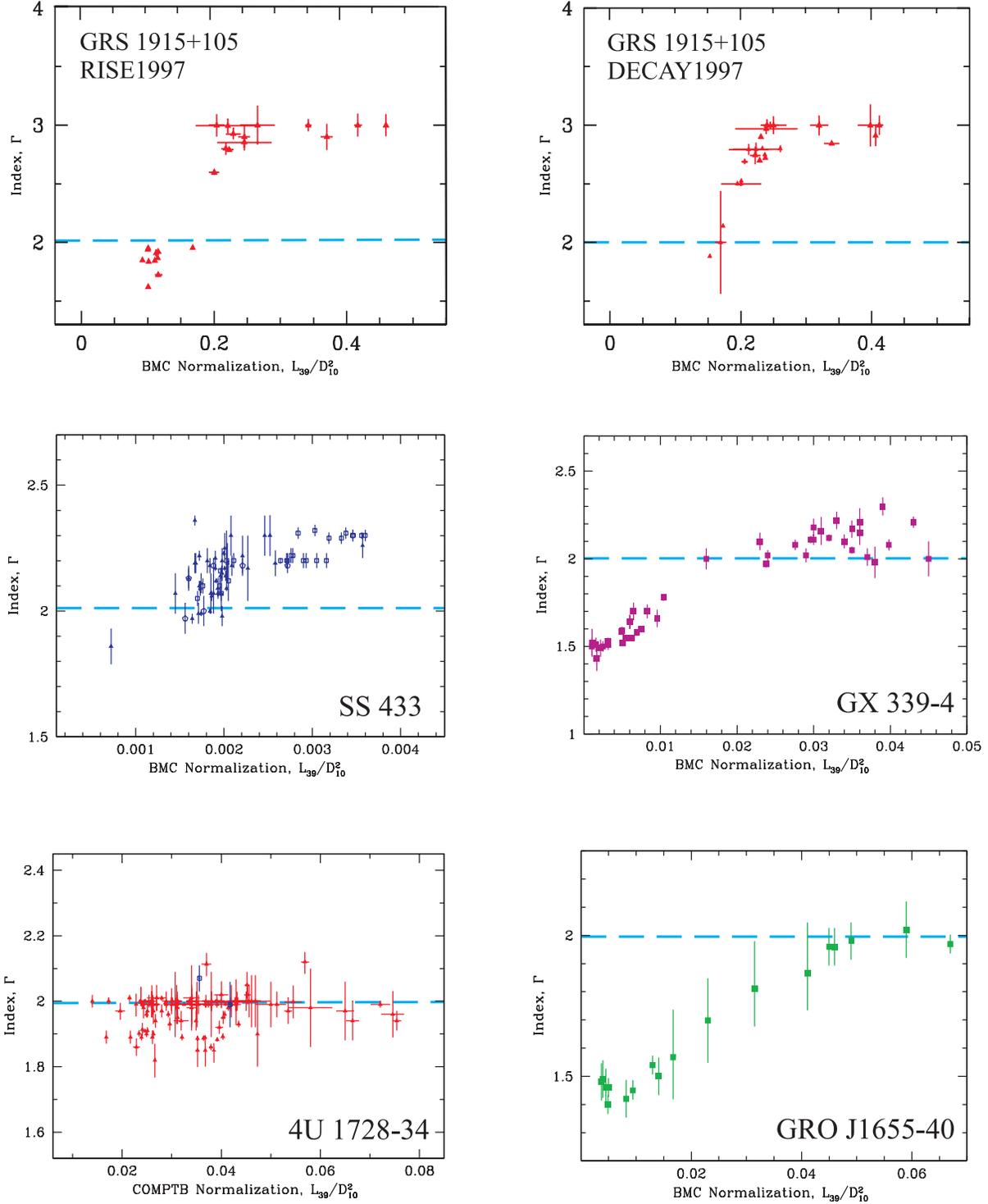}
\caption{Examples of diagrams of photon index $\Gamma$ versus mass accretion rate for BH candidate sources  
[GRS~1915+105 ({\it taken from} TS09), GX~339-4 (ST08), SS~433 (ST10) and GRO~J1655-40 (ST08)] along  with that for  $atoll$ 
NS  4U~1728-34. One can see a noticeable change of $\Gamma$ followed by saturation plateau for BHs as for NS 4U~1728-34 the index slightly varies about 2 (see also Fig. \ref{hist}).
% demonstrate different samples (patterns) of burst track: significant ch of photon 
%index $\Gamma$, accompanied with smooth (ordered correlated) growth and saturation section, for BH sources 
%and quazi-constantcy of $\Gamma$ for $atoll$ NS system. 
The level for $\Gamma=2$ is indicated by $blue$ dashed line.
%By $blue$ dashed line is indicated the $\Gamma$-level of 2.
}
\label{bh_ns_examples}
\end{figure}

\end{document}